\documentclass[prd ,aps, reprint, 10pt,nofootinbib,groupedaddress,preprintnumbers]{revtex4-1}
\pdfoutput=1
\usepackage{geometry}
\usepackage{graphicx}
\usepackage{amssymb}
\usepackage{mathtools}
\usepackage{epstopdf}
\usepackage{color}
\usepackage{bbold}
\usepackage{fullpage}
\usepackage{float}
\usepackage[title]{appendix}
\usepackage[colorlinks, citecolor=blue, linkcolor=black, urlcolor=purple]{hyperref}
\usepackage{amsfonts}
\usepackage{amsmath}
\usepackage{setspace}
\usepackage{epsfig}
\usepackage[caption=false]{subfig}
\usepackage[usenames, dvipsnames]{xcolor}
\usepackage{tabularx}
\usepackage{empheq}
\usepackage{mathrsfs}

\usepackage{units} % dimensionful quantity formatting
\newcommand{\eps}{\epsilon}
\newcommand{\order}[1]{\mathcal{O}{(#1)}}

\newcommand{\be}{\begin{eqnarray}}
\newcommand{\ee}{\end{eqnarray}}
\newcommand{\nn}{\nonumber}
\newcommand{\Hz}{\text{ Hz}}
\newcommand{\Lagr}{\mathcal{L}}
\newcommand{\beq}{\begin{equation}}
\newcommand{\eeq}{\end{equation}}

\newcommand{\wl}{\omega_\text{0}}
\newcommand{\ws}{\omega_\text{1}}

\newcommand{\gyy}{g_{a\gamma\gamma}}
\newcommand{\Bvec}{\mathbf{B}}
\newcommand{\Dvec}{\mathbf{D}}
\newcommand{\Evec}{\mathbf{E}}

\newcommand{\rhodm}{\rho_{_{\text{DM}}}}
\newcommand{\p}{\varphi}
\newcommand{\w}{\omega}

\newcommand{\me}{m_{\rm e}}
\newcommand{\xv}{{\bf x}}
\newcommand{\uv}{\boldsymbol{u}}
\newcommand{\xiv}{\boldsymbol{\xi}}
\newcommand{\Sv}{{\bf S}}
\newcommand{\eq}[1]{Eq.~(\ref{#1})}
\newcommand{\mamin}{10^{-14}}

\definecolor{darkblue}{rgb}{0.2,0.2,0.9}
\definecolor{colorRTD}{rgb}{.2,.2,.7}

\begin{document}

\title{Axion Dark Matter Detection by Superconducting Resonant Frequency Conversion}

\author{Asher~Berlin}
\affiliation{Center for Cosmology and Particle Physics, Department of Physics,
New York University, New York, NY 10003, USA.}
\author{Raffaele~Tito~D'Agnolo}
\affiliation{Institut de Physique Th\'eorique, Universit\'e Paris Saclay, CEA, F-91191 Gif-sur-Yvette, France}
\author{Sebastian~A.~R.~Ellis}
\author{Christopher~Nantista}
\author{Jeffrey~Neilson}
\author{Philip~Schuster}
\author{Sami~Tantawi}
\author{Natalia~Toro}
\author{Kevin~Zhou}
\affiliation{SLAC National Accelerator Laboratory, 2575 Sand Hill Road, Menlo Park, CA 94025, USA}

\begin{abstract}
We propose an approach to search for axion dark matter with a specially designed superconducting radio frequency cavity, targeting axions with masses $m_a \lesssim 10^{-6} \text{ eV}$. Our approach exploits axion-induced transitions between nearly degenerate resonant modes of frequency $\sim \text{GHz}$. A scan over axion mass is achieved by varying the frequency splitting between the two modes. Compared to traditional approaches, this allows for parametrically enhanced signal power for axions lighter than a GHz. The projected sensitivity covers unexplored parameter space for QCD axion dark matter for $10^{-8} \text{ eV} \lesssim m_a \lesssim10^{-6} \text{ eV}$ and axion-like particle dark matter as light as $m_a \sim 10^{-14} \text{ eV}$.
\end{abstract}

\maketitle
\onecolumngrid
%\tableofcontents

\section{Introduction}
%!TEX root = axion_arxiv.tex

The axion is a hypothetical parity-odd real scalar, protected by a shift symmetry and derivatively coupled to Standard Model fields. It is predicted by the Peccei--Quinn solution to the strong CP problem~\cite{Peccei:1977hh,Peccei:1977ur,Weinberg:1977ma,Wilczek:1977pj} and expected to arise generically from string theory compactifications~\cite{Svrcek:2006yi, Arvanitaki:2009fg, Stott:2017hvl}. It was shown to be a viable dark matter (DM) candidate four decades ago~\cite{Preskill:1982cy,Abbott:1982af}. A generic prediction of axion models is the coupling to photons~\cite{Dine:1982ah, Dine:1981rt,Zhitnitsky:1980tq,Kim:1979if,Shifman:1979if},
\be
\Lagr \supset - \frac{\gyy}{4}\, a\, F \tilde{F} = - \gyy \, a\, \mathbf{E}\cdot \mathbf{B}\, . \label{eq:photon}
\ee
This interaction can induce axion-photon conversion in the presence of a background electromagnetic field via the Primakoff process~\cite{Sikivie:1983ip}, which has been exploited in various axion searches~\cite{Boutan:2018uoc,Du:2018uak, Brubaker:2016ktl,PhysRevLett.59.839,Wuensch:1989sa,Hagmann:1990tj, Zhong:2018rsr,Blout:2000uc, PhysRevD.88.102003, PhysRevLett.116.161101, Anastassopoulos:2017ftl}. These searches have started to cover parameter space motivated by the Peccei--Quinn solution to the strong CP problem~\cite{Dine:1982ah, Dine:1981rt,Zhitnitsky:1980tq,Kim:1979if,Shifman:1979if}, $\gyy \simeq \unit[3 \times 10^{-16}]{GeV^{-1}} \, (m_a / \mu \text{eV})$, but for now without a positive detection.\footnote{The value quoted is the average of the DFSZ~\cite{Dine:1982ah, Dine:1981rt} and KSVZ~\cite{Zhitnitsky:1980tq,Kim:1979if,Shifman:1979if} predictions.}

More generally, an attractive motivation for axion-like particles (axions that do not solve the strong CP problem) is that they are a simple DM candidate. A very light axion can acquire a cosmological abundance from the misalignment mechanism that is in agreement with the observed DM energy density if $\gyy \sim 10^{-16} \text{ GeV}^{-1} ( m_a/\mu \text{eV} )^{1/4}$, where we have taken $\gyy \sim \alpha_\text{em} / 2 \pi f_a$ and assumed an $\order{1}$ initial misalignment angle (see Ref.~\cite{Blinov:2019rhb} for a recent discussion). This relation thus provides a cosmologically motivated target for axion-like particle searches.  

Cold axion DM produced by any mechanism generically virializes in the galactic halo. The typical virial velocity dispersion $v_a \sim 10^{-3}$ leads to an \textit{effective} quality factor of $Q_a \sim 1/\langle v_a^2\rangle \sim 10^6$. For timescales shorter than the axion coherence time $\tau_a \sim Q_a / m_a$, we can thus treat the axion as a monochromatic field of the form
\be
a(t)=\frac{\sqrt{2 \rhodm}}{m_a} \, \cos{m_a t} \, , \label{eq:axionTimeDep}
\ee
where $\rhodm \simeq 0.4 \text{ GeV}/ \text{cm}^3$ is the local DM energy density. Detailed coherence properties of the axion field have been discussed in Refs.~\cite{Sikivie:2009qn, Davidson:2014hfa}, but do not change the features noted above.

Resonant detectors are well-suited to exploit the coherence of the axion field. To date, most axion search experiments have matched the resonant frequency of the experiment to the mass of the axion DM being searched for. For $m_a \sim \mu\text{eV}$, the axion oscillates at $\sim$ GHz frequencies. This enables resonant searches using high-$Q$ normal-conducting cavities in static magnetic fields~\cite{Boutan:2018uoc,Du:2018uak,Brubaker:2016ktl,PhysRevLett.59.839,Wuensch:1989sa,Hagmann:1990tj,Zhong:2018rsr}, where a cavity mode is rung up through the interaction of Eq.~(\ref{eq:photon}), sourced by the axion field and the external $B$ field. These experiments take advantage of strong magnetic fields, the large quality factors achievable in GHz normal-conducting cavities, and low-noise readout electronics operating at the GHz scale. However, extending this approach to smaller axion masses would require the use of prohibitively large cavities. To probe lighter axions, experiments have been proposed using systems whose resonant frequencies are not directly tied to their size, such as lumped-element LC circuits~\cite{Sikivie:2013laa,Kahn:2016aff,Chaudhuri:2019ntz} or nuclear magnetic resonance~\cite{Budker:2013hfa}.

\begin{figure}[t]
\centering
\subfloat[][ Cartoon of cavity setup.]{
\hspace{-10mm}  \includegraphics[width = 0.35\columnwidth]{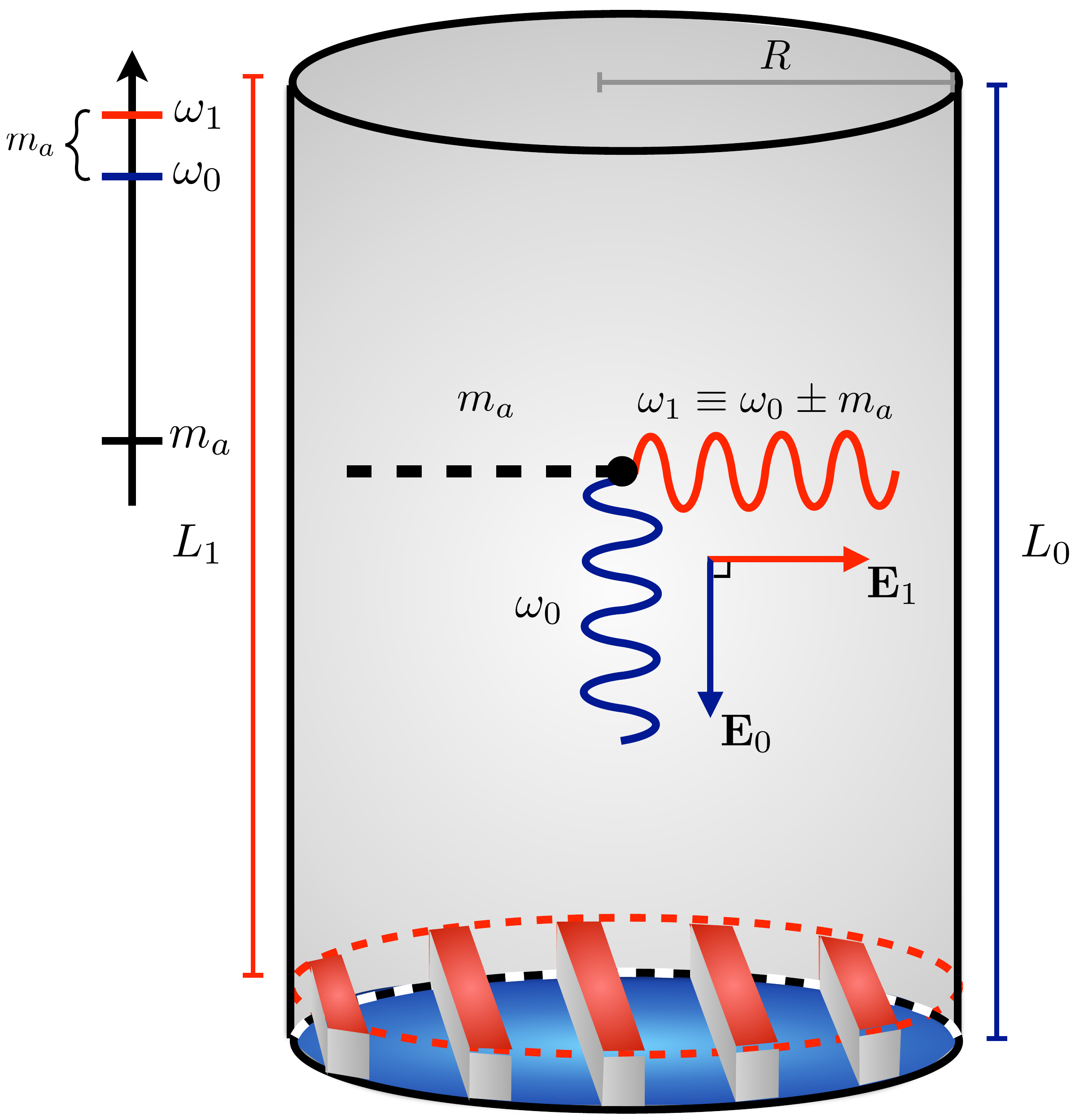}
}
\subfloat[][Signal parametrics.]{
\includegraphics[width = 0.6\columnwidth]{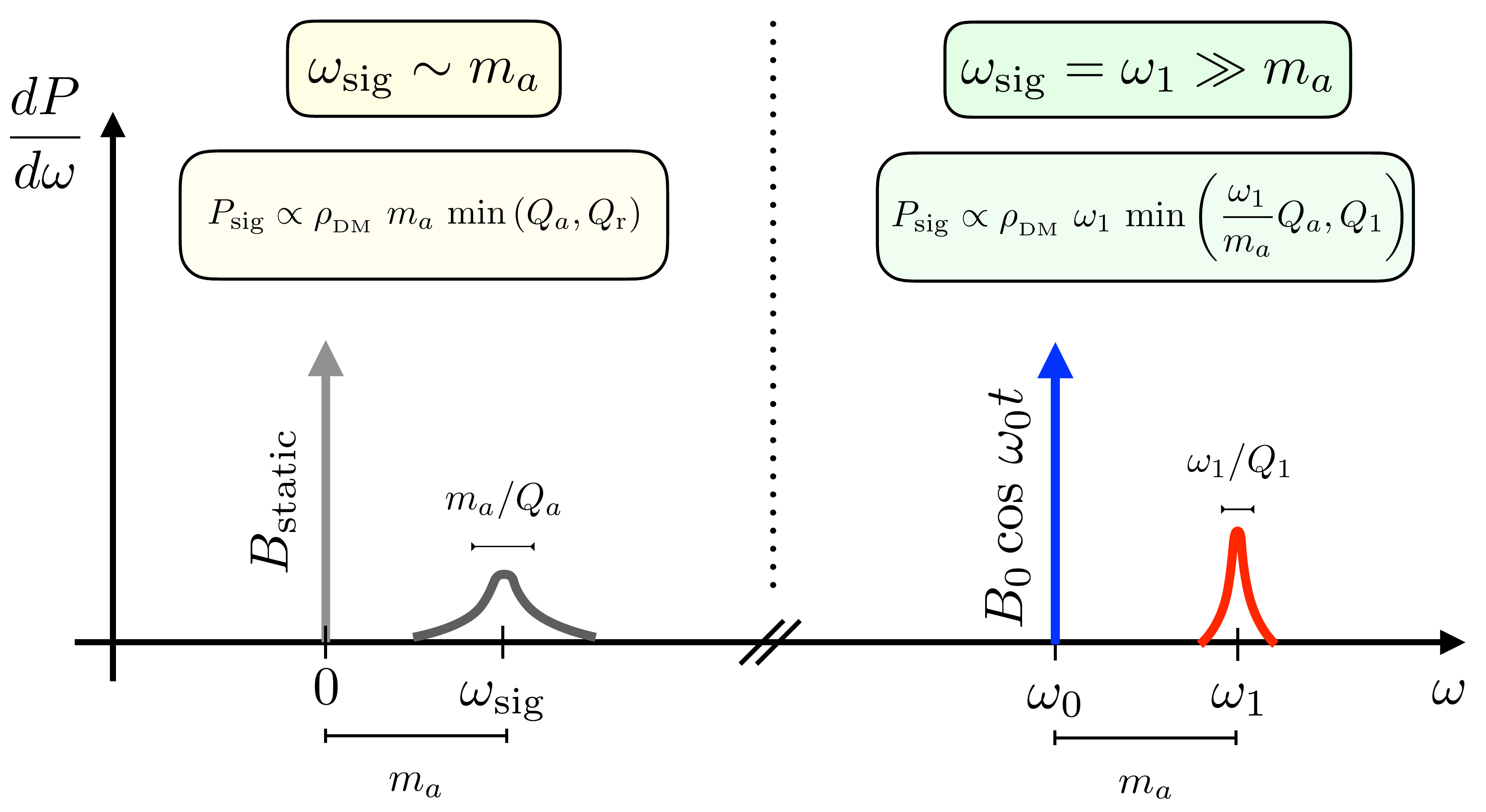}
}
\caption{\textbf{(a)} A schematic depiction of a potential cavity setup. A photon of frequency $\wl$ is converted by the axion dark matter background into a photon of frequency $\wl \pm m_a$, where $m_a$ is the axion mass. The cavity is designed to have two nearly degenerate resonant modes at $\wl$ and $\ws = \wl + m_a$. One possibility, as discussed in Section~\ref{sec:realisticCavity}, is to split the frequencies of the two polarizations of a hybrid $\text{HE}_{11p}$ mode in a corrugated cylindrical cavity. These two polarizations effectively see distinct cavity lengths, $L_0$ and $L_1$, allowing $\wl$ and $\ws$ to be tuned independently. In this case, larger frequency steps could be achieved by adjusting the fins (shown in red), while smaller frequency steps could be achieved with piezo-actuator tuners. \\
\textbf{(b)} A schematic comparison between the proposed frequency conversion scheme (right of the dotted line) and typical searches using static magnetic fields (left of the dotted line). The vertical and horizontal axes correspond to differential power and frequency, respectively, of either the driven field (vertical arrows) or the axion-induced signal (resonant curves). The parametric signal power derived in Section~\ref{sec:overview} is shown for both setups, where we assume $\omega_{\text{sig}} \sim V^{-1/3}$ for our proposed scheme and factored out a common volume dependence of $V^{5/3}$.
} 
\label{fig:reach}
\end{figure}

\begin{figure}[t]
\centering
\includegraphics[scale=0.95]{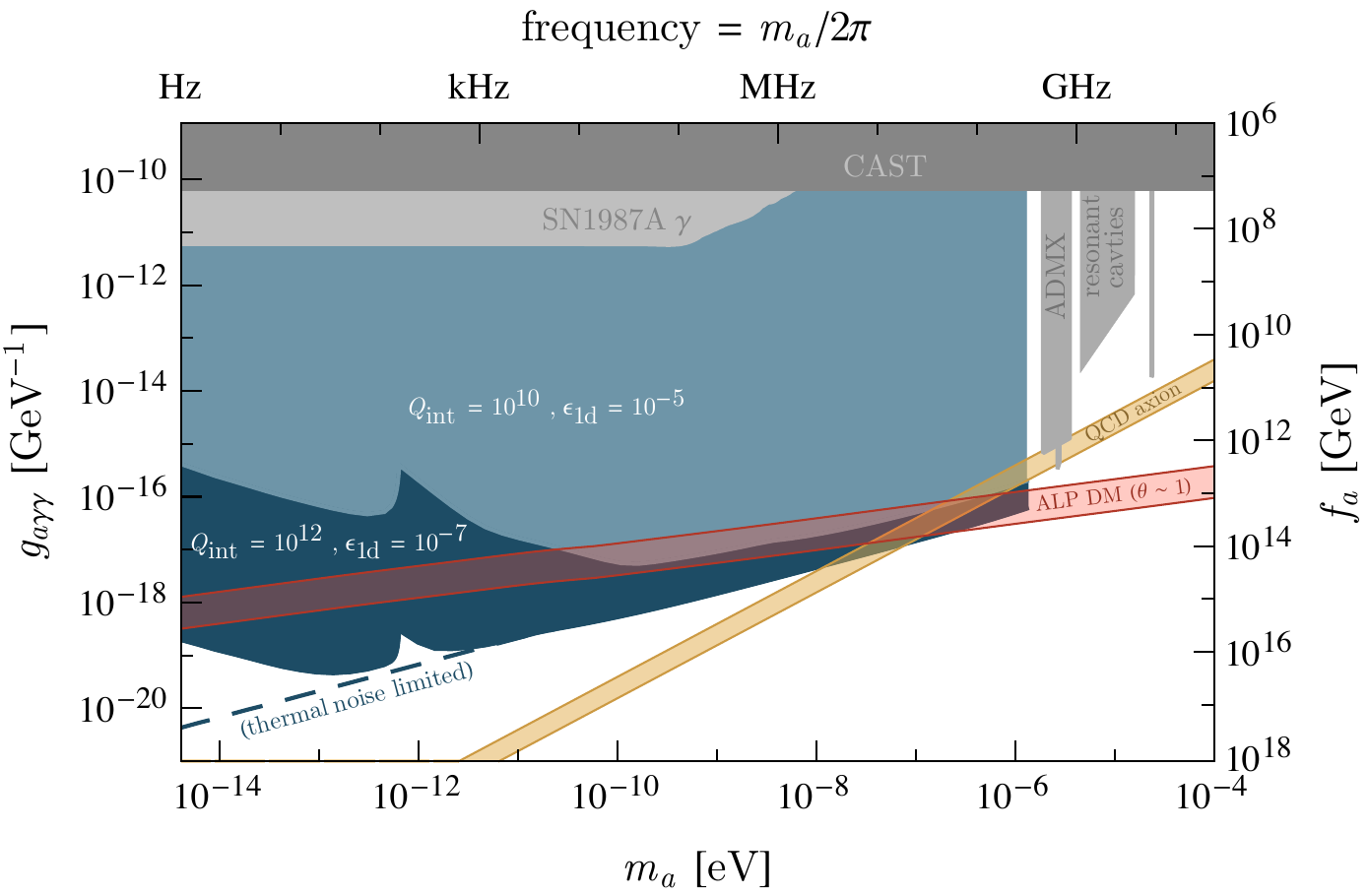}
\caption{The anticipated reach to axion dark matter in the $\gyy-m_a$ plane, for various experimental configurations, compared to existing constraints, shown in gray. Along the right axis, we relate the axion-photon coupling to the symmetry breaking scale $f_a$ by $\gyy \sim \alpha_\text{em} / 2 \pi f_a$. As two representative examples, we show the projected sensitivity assuming an intrinsic quality factor and readout-pump mode coupling (see Section~\ref{sec:leakage}) of $Q_\text{int} = 10^{10}$, $10^{12}$ and $\eps_{1\text{d}} = 10^{-5}$,  $10^{-7}$, respectively. The dashed line shows the thermal noise limited sensitivity for $Q_\text{int} = 10^{12}$ and $\eps_{1\text{d}} = 10^{-7}$. In all cases, we assume a pump mode frequency of $\omega_0 / 2 \pi = \text{GHz}$, a cavity volume of $V = \unit[1]{m^3}$, a peak magnetic field of $B_0 = 0.2 \text{ T}$, a mode overlap of $\eta_{10} = 1$ (see Eq.~(\ref{eq:overlap})), a cavity temperature of $T = \unit[1.8]{K}$, an average wall displacement of $q_\text{rms} = \unit[10^{-1}]{nm}$ (as defined in Section~\ref{sec:noiseVibration}), and an $e$-fold time of $t_{e} = \unit[10^7]{s}$. The orange band denotes the range of couplings and masses as motivated by the strong CP problem. Along the red band, axion production through the misalignment mechanism is consistent with the observed dark matter energy density, assuming an $\order{1}$ initial misalignment angle. As discussed in Section~\ref{sec:noiseVibration}, the feature near $m_a \sim \text{kHz}$ is due to our assumption that there are no mechanical resonances below a kHz.
}
\label{fig:genreach}
\end{figure}

In this work, we explore an alternative approach to resonant axion detection, where the frequency \emph{difference} between two modes is tuned to be on-resonance with the axion field, while the mode frequencies themselves remain parametrically larger. Because of their large quality factors, superconducting radio frequency (SRF) cavities are ideal resonators for such a setup. More concretely, as illustrated in Figure~\ref{fig:reach}, we consider an SRF cavity with a small, tunable frequency difference between two low-lying modes, which we call the ``pump mode'' and the ``signal mode.'' The cavity is prepared by driving the pump mode, which has frequency $\w_0 \sim \text{GHz} \gg m_a$. If the signal mode is tuned to a frequency $\w_1 \simeq \w_0 \pm m_a$, then the axion DM field resonantly drives power from the pump mode to the signal mode.

The idea of detecting axions through photon frequency conversion has been studied in other contexts.\footnote{ Different SRF setups have also been considered for production and detection of light, non-DM axions~\cite{PhysRevLett.123.021801,PhysRevD.100.015036}. Another, distinct idea is the proposal of Refs.~\cite{Goryachev:2018vjt,Thomson:2019aht} to drive two modes and detect the resulting axion-induced frequency shifts.} These include axion detection with optical cavities~\cite{Melissinos:2008vn,liu2019searching,Martynov:2019azm} and frequency conversion in SRF cavities with GHz-scale mode splittings~\cite{Sikivie:2010fa}. More generally, frequency conversion is a commonly used technique in signal processing, under the name of ``heterodyne detection."

However, frequency conversion in SRF cavities is particularly powerful because of the combination of high $Q$-factors and the large amount of stored energy in the pump mode. In this work, we highlight the parametric advantages of this approach at low axion masses, discuss scenarios for realizing the mode overlap and tunability requirements for such an experiment, and analyze key sources of noise. In the latter two aspects, we benefit from the decades-long effort to detect kHz-to-MHz gravitational waves with SRF cavity resonators~\cite{Pegoraro_1978}. The results from the prototypes of Refs.~\cite{Reece:1984gv,Bernard:2001kp,Bernard:2002ci,Ballantini:2005am} are particularly useful in anticipating the experimental challenges of our proposed approach.

Our study shows that axion-induced frequency conversion in SRF cavities could be sensitive to QCD axions for $\unit[10^{-8}]{eV} \lesssim m_a \lesssim \unit[10^{-6}]{eV}$ and axion DM as light as $m_a \sim \mamin\;{\rm eV}$. The projected sensitivity for two representative sets of experimental parameters is shown in Figure~\ref{fig:genreach}, with a larger set of parameters shown in Figure~\ref{fig:four_plot}. Compared to traditional resonant searches, fixing the signal to GHz frequencies leads to several advantages for lower axion masses:
\begin{enumerate}
\item High frequency readout leverages the large quality factors of SRF cavities, which are typically of order $Q \gtrsim 10^{10}$. In this case, the signal power saturates once $Q \gtrsim (\text{GHz} / m_a) \, Q_a$, unlike static-field detectors whose signal power saturates once $Q \gtrsim Q_a$.
\item Only a small fraction of the signal power ($m_a / \text{GHz} \ll 1$) is sourced directly by the axion DM field. Therefore, the signal is not suppressed by the small axion mass when its Compton wavelength is much larger than the detecting apparatus. This is unlike static-field electromagnetic resonators, where the signal power scales as $m_a$ in this limit. 
\item Operating readout electronics near the standard quantum limit has been demonstrated at GHz frequencies~\cite{Brubaker:2016ktl}. 
\end{enumerate}

In the next section, we present a parametric estimate of the axion-induced signal power and compare it to that of other resonant setups. In Section~\ref{sec:signal}, we provide a more detailed calculation, using a simple model without explicit reference to cavity parameters. We discuss a more complete experimental setup in Section~\ref{sec:realisticCavity}, deferring a detailed discussion of SRF cavity geometries to Appendix~\ref{sec:app}. In Section~\ref{sec:noise}, we study the expected sources of noise, with additional details in Appendix~\ref{app:rejection}. In Section~\ref{sec:reach}, we estimate the physics reach, with further detail regarding optimization of the readout coupling presented in Appendix~\ref{sec:overcouple}. Finally, we conclude in Section~\ref{sec:outlook}.

\section{Conceptual Overview} \label{sec:overview}
%!TEX root = axion_arxiv.tex

At the level of Maxwell's equations, an oscillating axion DM field sources a time-dependent effective current density, $J_\text{eff}$, in the presence of an applied magnetic field $B_0(t)$, of magnitude 
\be
J_\text{eff} (t) \sim \gyy \, B_0(t) \, \sqrt{\rhodm} \, \cos{m_a t}
\, .
\ee
This effective current density leads to a real magnetic field, $B_a \propto J_\text{eff}$. The oscillations of this field generate a small electromotive force
\be
\label{eq:emf1}
\mathcal{E}_a \sim V^{2/3} \, \partial_t B_a
\, ,
\ee
which can drive power into a resonant detector of volume $V$. In typical setups, the applied magnetic field is static, such that $\mathcal{E}_a^{(\text{static})} \propto m_a$. In the approach we advocate for here, the applied magnetic field oscillates in time, $B_0(t) = B_0 \cos \wl t$. Compared to static-field detectors of comparable size, the electromotive force is significantly larger,
\be
\label{eq:conceptualemf}
\frac{\mathcal{E}^{(\text{osc.})}_a}{\mathcal{E}^{(\text{static})}_a} \sim \frac{\wl + m_a}{m_a} \sim \frac{\w_1}{m_a}
\, .
\ee
This is the essential reason for the parametric enhancement of our approach at low axion masses ($m_a \ll \w_0$).\footnote{There is a well-known argument that axion signals must degrade at small $m_a$, since the massless limit at fixed axion field amplitude would be equivalent to a static QED $\theta$-angle. The scaling of Eq.~(\ref{eq:emf1}) does not violate this argument because $J_{\text{eff}}\propto \sqrt{\rhodm}\sim m_a \, a$. Thus, for a fixed axion field amplitude, the electromotive force in our setup scales as $m_a$, compared to $m_a^2$ for static-field experiments.}

To make this intuition more precise, it is useful to compute the signal power explicitly and compare it to that of static-field resonators. In general, the power delivered to a resonator of volume $V$ and resistance $R$ is
\be
\label{eq:conceptualpower}
P_\text{sig}^{(\text{r})} \sim \frac{\mathcal{E}_a^2}{R} ~ \text{min} \Big( 1 , \frac{\tau_a}{\tau_\text{r}} \Big) \sim\w_\text{sig}^2 \, B_a^2 V \, \text{min}(Q_\text{r} / \w_\text{sig}, Q_a / m_a)
\, ,
\ee
where $\tau_\text{r} \sim Q_\text{r} / \w_\text{sig}$ is the ring-up time for a resonator with quality factor $Q_\text{r}$ and readout frequency $\w_\text{sig}$, and in the second equality, we expressed $R$ in terms of $Q_\text{r}$. Note that as a function of $Q_\text{r}$, the signal power saturates once the axion coherence time is smaller than the resonator ring-up time, since only a fraction of the axion power resides within the resonator bandwidth, as encapsulated in the second factor in both equalities.
 
To date, most resonant experiments searching for electromagnetically coupled axion DM employ static magnetic fields, since these are more easily sourced at large field strengths. In this case, $J_\text{eff} (t) \propto \cos{m_a t}$ implies that this current density sources photons of energy and frequency comparable to $m_a$, which can be detected with an apparatus whose resonant frequency is matched to the axion rest mass. For $m_a \sim \text{GHz}$, this is the strategy employed by resonant cavity experiments such as ADMX~\cite{Boutan:2018uoc, Du:2018uak}. However, for any static-field cavity detector, this approach becomes increasingly difficult for $m_a \ll \text{GHz}$, since the resonant frequency is typically controlled by the inverse length-scale of the apparatus. 

By contrast, LC resonators can search for sub-GHz axions because their resonant frequency is not directly tied to the geometric size of their circuit components. In such a setup, when the Compton wavelength of the axion is much larger than the shielded detection region of volume $V$, the size of the axion-induced magnetic field follows simply from the quasistatic expectation, $B_a \sim J_\text{eff} \, V^{1/3}$. Since the readout frequency of static-field setups is dictated by the axion mass, the signal power of an LC circuit with quality factor $Q_{\text{LC}}$ is parametrically
\be \label{eq:heuristic_lc_signal_power}
P_\text{sig}^{(\text{LC})} \sim  m_a \, J_\text{eff}^2 \, V^{5/3} \, \text{min}(Q_{\text{LC}}, Q_a)
\, .
\ee
The saturation of signal power at $Q_\text{r} \gtrsim Q_a$ as well as the overall suppression at small axion masses is characteristic of static-field setups. This latter point can also be understood from the fact that for a static-field configuration, the axion-induced electromotive force vanishes for zero axion mass and fixed DM energy density, since $\mathcal{E}_a \propto m_a$.

Our setup instead involves driving a resonant cavity at a frequency $\w_0 \gg m_a$. An axion DM background converts the frequency, sourcing an effective current oscillating at $\w_{\text{sig}} = \w_1 = \w_0 \pm m_a$,
\be
J_\text{eff} (t) \sim \gyy \, B_0 \, \sqrt{\rhodm} \, \cos{(\w_0 \pm m_a ) t}
\, ,
\ee
which drives power into the signal mode. In this case, $B_a \sim J_\text{eff} / \w_1$, and for a fixed DM energy density, the electromotive force is not suppressed for $m_a \ll \text{GHz}$ since $\mathcal{E}_a \propto \w_1$. By the same logic as the previous calculation, the axion-induced signal power is
\be \label{eq:heuristic_signal_power}
P_\text{sig} \sim J_\text{eff}^2 \, V \, \text{min} (Q_\text{r} / \w_1 , Q_a/m_a)
\, ,
\ee
which yields a parametric advantage\footnote{Axion detection by frequency conversion in a radio frequency cavity was also briefly considered in Ref.~\cite{Goryachev:2018vjt}, but the authors did not find the same parametric enhancement we demonstrate here.} over LC resonators when $m_a \ll V^{-1/3}$. Intuitively, this is because each axion-photon interaction in the cavity involves a photon of energy $\w_0$, and so only a small fraction ($m_a / \w_0 \ll 1$) of this signal power is contributed by the axion background, with the remainder originating from the pump mode. Since $\w_1 \gg m_a$, maximizing the signal power in our setup requires resonator quality factors much larger than $Q_a$, saturating only when $Q_\text{r} \gtrsim (\w_1 / m_a) \, Q_a \gg 10^6$. This motivates the choice of using an SRF cavity, as superconducting resonators have been built with quality factors as large as $Q_\text{r} \sim \text{few} \times 10^{11}$.

To complete our overview, we give a parametric comparison of the reach. This can be done straightforwardly when our approach is thermal noise limited, which occurs in the right half of Figure~\ref{fig:genreach}. As we will see in Section~\ref{sec:reach}, accounting for the scan rate and coupling optimization leads to simple expressions for the signal-to-noise ratio (SNR), which do not require the casework of Eqs.~\eqref{eq:heuristic_lc_signal_power} and~\eqref{eq:heuristic_signal_power}. Instead, for general quality factors,
\be
\label{eq:SNR_comparison}
\frac{\text{SNR}}{\text{SNR}^{(\text{LC})}} \sim \frac{\w_1}{m_a} \left( \frac{Q_{\text{int}}}{Q_{\text{LC}}} \right)^{1/2} \left( \frac{T_{\text{LC}}}{T} \right)^{1/2} \left( \frac{B_0}{B_{\text{LC}}} \right)^2 \, ,
\ee
where we took $\w_1 \sim V^{-1/3}$, which holds for low-lying cavity modes, and $Q_{\text{int}}$ is the intrinsic quality factor of the SRF cavity. For the reference parameters of the lower curve of Figure~\ref{fig:genreach} and comparison parameters $Q_{\text{LC}} = Q_a \sim 10^6$, $T_{\text{LC}} = \unit[0.1]{K}$, and $B_{\text{LC}} = \unit[4]{T}$, the last three factors roughly cancel, leaving an enhancement factor of $\omega_1/m_a$. In the next section, we begin the work of establishing these results, by directly solving the relevant equations of motion to compute the signal power.

\section{Signal Power} \label{sec:signal}
%!TEX root = axion_arxiv.tex

In this section, we explicitly compute the signal induced by axion DM interacting with a loaded cavity. For this calculation, it suffices to use a simplified model that treats the cavity as a collection of fixed cavity modes. In the following two sections, we refine this model by including the additional layers of complexity needed to describe the system in the presence of noise. 

Our starting point is Maxwell's equations modified by the axion interaction of Eq.~\eqref{eq:photon},
\be \label{eq:maxwell}
\nabla \cdot \Evec &=& \rho - \gyy \Bvec \cdot \nabla a \ , \nn \\
\nabla \times \Bvec &=& \partial_t \Evec + \mathbf{J} - \gyy \left( \Evec \times \nabla a - \Bvec \, \partial_t a \right)\, . 
\ee
Since the spatial gradients of the axion field are small, the dominant effect is that the axion sources an effective current, $\mathbf{J}_\text{eff} = \gyy \Bvec \partial_t a$. The effective current $\mathbf{J}_\text{eff}$ inherits its time-dependence from the oscillating axion and pump mode magnetic field and can resonantly drive power into other cavity modes with matching characteristic frequency. 

To compute the steady state signal power, it is convenient to work in frequency space. In doing so, we adopt the following convention for the Fourier transform of a function $f$,
\begin{align}
f(t) = \frac{1}{2 \pi} \, \int d \w ~ e^{i \w t} f( \w) \, , \quad f(\w) = \int dt ~ e^{-i \w t} f(t) \, .\nn
\end{align}
When unspecified, the region of integration for $\w$ or $t$ is implicitly $-\infty$ to $\infty$. We define the power spectral density (PSD) of $f$, denoted as $S_f(\w)$, by
\be
\label{eq:PSDdef}
\langle f(\w) f^*(\w^\prime) \rangle = S_f(\w) \, \delta (\w - \w^\prime)
~,
\ee
so that the steady state average power can be expressed as\footnote{In Eq.~(\ref{eq:PSDdef}), the brackets denote an ensemble average, where a signal $f(t)$ is Fourier transformed in many different time intervals, which are then averaged in a given frequency bin. Eq.~(\ref{eq:avgPSD}) then defines $\langle f(t)^2 \rangle$, which can equivalently be described as a time average of $f(t)^2$.}
\be
\label{eq:avgPSD}
\langle \, f (t)^2 \, \rangle = \frac{1}{ (2 \pi)^2} \, \int d \w ~ S_f(\w)
~.
\ee
Note that all PSDs in this work are two-sided.

Given the tiny backreaction of the axion field on the cavity, it is useful to decompose the electric and magnetic fields into a set of vacuum cavity modes:
\be
\label{eq:modedecomposition}
\Evec(t, \mathbf{r}) &=&  \sum_n {\Evec}_n(t, \mathbf{r})=\sum_n e_n (t) \, \tilde{\Evec}_n(\mathbf{r})  \ \nn , \\ \Bvec(t, \mathbf{r}) &=& \sum_n {\Bvec}_n(t, \mathbf{r})=\sum_n b_n (t) \, \tilde{\Bvec}_n(\mathbf{r})  \ ,
\ee
where the resonant modes satisfy the conditions
\be
\nabla^2 \tilde{\Evec}_n = - \w_n^2 \, \tilde{\Evec}_n \ , &\quad& \nabla^2 \tilde{\Bvec}_n = - {\w}_n^2 \, \tilde{\Bvec}_n\, , \nn\\
\int_V \tilde{\Evec}_n^* \cdot \tilde{\Evec}_l =  \delta_{nl}\int_V |\tilde \Evec_n|^2, &\quad& \int_V \tilde{\Bvec}_n^*\cdot \tilde{\Bvec}_l = \delta_{nl}\int_V |\tilde \Bvec_n|^2 \, .
\ee
Here, $V$ is the volume of the cavity and $\w_n$ are the resonant frequencies. Using the above definitions, Maxwell's equations in Eq.~\eqref{eq:maxwell} can be rewritten as an equation of motion for the cavity's electric field in the presence of background axion and magnetic fields,
\be 
\sum_n \left( \w^2 - \w_n^2 - i \, \frac{\w \, \w_n}{Q_n}\right){\Evec}_n (\w)= \gyy \int dt ~ e^{-i \w t} ~ \partial_t (\Bvec \, \partial_t a) 
 \label{eq:signal1}\, ,
\ee
where we have neglected terms proportional to the small axion velocity. Each mode has a distinct quality factor, $Q_n$, that is dictated by the electric field profile near the walls and power losses through the loading and readout ports, and determines the dissipative terms on the left-hand side. Above, we have neglected the motion of the cavity walls, which can couple distinct modes and shift their resonant frequencies; we account for this source of noise in Section~\ref{sec:noiseVibration}. 

To complete the calculation, we note that the magnetic field in Eq.~\eqref{eq:signal1} is dominated by the pump mode such that $\Bvec \simeq \Bvec_0$. We then define the characteristic amplitude of the pump mode magnetic field as
\be
B_0 \equiv \sqrt{\frac{1}{V} \int_V |\tilde \Bvec_0|^2}
~.
\ee
The steady state average power delivered to the signal mode ($n = 1$) can be written in terms of a signal PSD defined analogously to Eq.~\eqref{eq:avgPSD},
\be 
\label{eq:total_signal_power}
P_\text{sig} = \frac{\ws}{Q_1} \, U_1= \frac{1}{(2 \pi)^2} \int d\w ~ S_\text{sig} (\w)
~,
\ee
where $U_1$ is the electromagnetic energy stored in the signal mode. From Eq.~\eqref{eq:signal1}, we find that
\be
\label{eq:SignalPSD}
S_\text{sig} (\w) = \frac{\w_1}{Q_1} \, (\gyy \, \eta_{10}\, B_0)^2 \, V \, \frac{\w^2}{(\w^2 - \w_1^2)^2 + (\w \, \w_1 /Q_1)^2}  \int \frac{d\w^\prime}{(2\pi)^2} \, (\w^\prime-\w)^2 \, S_{b_0}(\w^\prime) \, S_a(\w-\w^\prime) 
\, ,
\ee
where $S_a (\w)$ is the axion PSD, $S_{b_0} (\w)$ is the PSD for $b_0 (t)$ (defined in Eq.~(\ref{eq:modedecomposition})), and $\eta_{10}$ is an $\order{1}$ mode overlap factor, 
\be
\label{eq:overlap}
\eta_{10}\equiv \frac{ \left| \int_V \tilde \Evec^*_1 \cdot \tilde \Bvec_0 \right|}{\sqrt{\int_V |\tilde \Evec_1|^2} \; \sqrt{\int_V |\tilde \Bvec_0|^2}}\leq 1 
\, .
\ee
We have ignored backreaction on the axion field, as this is negligible even for very large quality factors. We note that Eq.~\eqref{eq:SignalPSD} is only valid when the experimental integration time $t_{\text{int}}$ exceeds both the ring-up time of the signal mode, $\tau_r \sim Q_1 / \w_1$, and the axion coherence time, $\tau_a \sim Q_a / m_a$. The steady state power is achieved when $t_{\text{int}} \gtrsim \tau_r$, but if $t_{\text{int}} \lesssim \tau_a$, the axion PSD is not resolved, and $S_a(\w)$ must be convolved with a window function.\footnote{Our result also breaks down in the extreme case $m_a \lesssim \w_1 / Q_1 \sim \unit[10^{-17}]{eV} \times (10^{12}/Q_1)$, where the axion oscillates on a longer timescale than the ring-up time. In this case, the signal power does not reach a steady value, but rather depends on the instantaneous phase of the axion field. Eq.~\eqref{eq:sigpower} remains valid only if $P_{\text{sig}}$ is taken to denote the average power over an entire axion field oscillation. This is not relevant for any of the parameter space shown in Figure~\ref{fig:genreach}. For the smaller intrinsic quality factors or $e$-fold times shown in Figure~\ref{fig:four_plot} (which affect $Q_1$, as described in Section~\ref{sec:reach}), we restrict our calculations to $m_a \gtrsim \w_1 / Q_1$.}  

If the spectral width of the pump mode magnetic field is sufficiently narrow, then it may be approximated as a monochromatic source, $b_0(t) = \cos{\w_0 t}$, which corresponds to
\be
S_{b_0}(\w)=\pi^2\left[\delta(\w-\w_0)+\delta(\w+\w_0)\right]
\, .
\ee
Eq.~\eqref{eq:SignalPSD} then reduces to
\be
\label{eq:SignalPSDnarrow}
S_\text{sig} (\w) = \frac{\w_1}{4 Q_1} \left(\gyy \, \eta_{10} \, B_0 \right)^2 V ~~ \frac{\w^2 \, \big[ (\w-\w_0)^2 \, S_a (\w-\w_0) + (\w+\w_0)^2 \, S_a (\w+\w_0)\big]}{(\w^2 - \w_1^2)^2 + (\w \, \w_1 / Q_1)^2}
\, .
\ee
As we will see in Section~\ref{sec:reach}, this is a valid approximation in most of the parameter space considered in this work. This is possible because the magnetic field can have a much narrower width than the pump mode itself, as its width is determined by the frequency stability of the oscillator that loads the cavity.

To understand Eq.~(\ref{eq:SignalPSDnarrow}) parametrically, we assume the signal mode frequency is on resonance and consider two limiting cases. The frequency spread of the axion PSD is controlled by its effective quality factor $Q_a \sim 10^6$. If the axion is narrow compared to the signal mode's bandwidth ($m_a/Q_a \ll \w_1/Q_1$), we can evaluate the integral of Eq.~(\ref{eq:total_signal_power}) by treating the axion PSD as a delta function. Instead, if the axion is broad compared to the signal bandwidth ($m_a / Q_a \gg \w_1 / Q_1$), we can evaluate the integral using the narrow width approximation for the Breit--Wigner response of the signal mode. The result is
\be
\label{eq:sigpower}
P_\text{sig} \simeq \frac{1}{4} \left( \gyy \, \eta_{10} \, B_0 \right)^2 \, \rhodm \, V \times \begin{cases} Q_1 / \w_1 & \frac{m_a}{Q_a} \ll \frac{\w_1}{Q_1} 
\\ 
\pi Q_a/m_a & \frac{m_a}{Q_a} \gg \frac{\w_1}{Q_1} \, , \end{cases} 
\ee
which matches the parametric estimate of Eq.~\eqref{eq:heuristic_signal_power}. Here we use the normalization
\be \label{eq:dm_normalization}
\langle a(t)^2 \rangle = \frac{1}{ (2 \pi)^2} \, \int d \w ~ S_a(\w) = \frac{\rhodm}{m_a^2}
\, ,
\ee
and take $S_a(\w)$ to be governed by a virialized Maxwellian velocity distribution~\cite{Krauss:1985ub}. 

For large axion masses, the axion is broad, and the signal power in Eq.~\eqref{eq:sigpower} is suppressed by $m_a^{-1}$ since only a small fraction of the axion PSD lies within the detector bandwidth. As the axion mass decreases, the signal power increases, saturating when these two bandwidths are comparable, i.e., when the axion coherence time matches the ring-up time of the signal mode, $\tau_a \sim \tau_r$. As discussed in the previous section, this differs from resonant experiments where the readout frequency is comparable to the axion mass, in which case the signal power saturates once $Q_1 \gtrsim Q_a$. 

Expanding on the intuition developed in Section~\ref{sec:overview}, we now compare more carefully the parametric form of the signal power in Eq.~(\ref{eq:sigpower}) to that of static-field experiments designed to resonantly detect axions with masses $m_a \ll \text{GHz}$. For example, near-future LC resonators plan on using magnetic fields of size $B_\text{LC} \sim 4 \text{ T}$, while the magnetic fields for our setup can be no larger than roughly $0.2 \text{ T}$, to preserve the superconducting properties of the cavity. However, this is compensated by the much larger quality factors attainable by SRF cavities. To see this, note that a static-field LC resonator is required to operate in the quasistatic limit once $m_a \ll V^{-1/3}$. In this case, as discussed in Section~\ref{sec:overview}, the parametric form of the signal power is
\be
\label{eq:PLC}
P_\text{sig}^\text{(LC)} \sim (\gyy B_\text{LC})^2 \, \rhodm \, V^{5/3} \, {\rm min}(Q_\text{LC}, Q_a) \, m_a 
\ .
\ee
The factor of $m_a$ in Eq.~\eqref{eq:PLC} stands in contrast to Eq.~\eqref{eq:sigpower}, and appears because the signal frequency in such an experiment is comparable to the axion mass. This is not the case for the setup discussed here because the signal frequency is always fixed to be $\w_0 + m_a \sim \w_1 \sim \text{GHz}$ even for $m_a \ll \text{GHz}$. Comparing Eqs.~(\ref{eq:sigpower}) and (\ref{eq:PLC}), we have
\begin{align}
\label{eq:LCcompare}
\frac{P_\text{sig}}{P_\text{sig}^\text{(LC)}} \sim \left( \frac{0.2 \text{ T}}{4 \text{ T}} \right)^2
\times
\begin{cases}
\left( Q_1 / Q_a \right)^2 \frac{(\w_1/Q_1)}{(m_a / Q_a)} & \frac{m_a}{Q_a} \ll \frac{\w_1}{Q_1}
\\
\left( \w_1 / m_a \right)^2 & \frac{m_a}{Q_a} \gg \frac{\w_1}{Q_1}
\, ,
\end{cases} 
\end{align}
where we took the cavity and LC resonator to be of comparable size, fixed $\w_1 V^{1/3} \sim 1$ for the cavity setup, and set $Q_{\text{LC}} \sim Q_a$. Eq.~(\ref{eq:LCcompare}) shows that a frequency conversion setup using an SRF cavity has a parametric advantage in signal power when $m_a \ll \w_1$, which is the regime shown in Figure~\ref{fig:genreach}. For a broad axion, $m_a \lesssim \w_1/20$ is already enough to overcome the weaker magnetic field, while for a narrow axion the larger quality factors achievable in SRF cavities more than suffice to compensate at any axion mass.

Of course, this does not suffice to establish a comparably enhanced sensitivity, since noise sources can vary drastically across different experimental setups. We investigate these noise sources in detail in Section~\ref{sec:noise}. Realistic values for the relevant cavity parameters are discussed in more detail in the next section.

\section{A Cavity Concept} \label{sec:realisticCavity}
%!TEX root = axion_arxiv.tex

In this section, we discuss the choice of cavity geometry and pump and signal modes, as well as the quality factors attainable in SRF cavities. We also outline possible methods for tuning the mode splitting $\omega_1 - \omega_0$, loading the cavity, and reading out the signal. 

As mentioned in Section~\ref{sec:signal}, the peak magnetic field in an SRF cavity will be smaller than in a conventional RF cavity, and this must be compensated by a larger quality factor. In multi-cell elliptical cavities operating at GHz frequencies designed for accelerating charged particle beams, intrinsic quality factors of $Q_{\rm int} \simeq 4 \times 10^{10}$ (and in one case as high as $Q_{\rm int} \gtrsim 2\times 10^{11}$) have been achieved~\cite{Romanenko:2014yaa, Posen:2018bjn}, a factor of over $10^6$ greater than what the same geometry would display in warm copper. However, we are not restricted to geometries useful for particle acceleration. Quality factors of $Q_{\text{int}} \sim 10^5$ are commonly achieved in overmoded non-superconducting RF cavities with non-accelerator geometries~\cite{Farkas:1974ur, Tantawi:2014nxa, Wang:2017cbs}. This suggests that SRF counterparts can be constructed with quality factors as large as $Q_{\rm int} \sim 10^{12}$.\footnote{The power dissipation of a cavity with the parameters of Fig.~\ref{fig:genreach} would be $P_{\rm in} \sim 10^4 \times (10^{10}/Q_{\rm int})$ W. As such, operating SRF cavities with intrinsic quality factors significantly lower than $10^{10}$ is not practical due to power and cooling demands.}

We now consider the choice of cavity geometry, where the goal is to find a cavity design with two nearly degenerate modes and an $\order{1}$ geometric overlap factor $\eta_{10}$, as defined in Eq.~\eqref{eq:overlap}. Rectangular, cylindrical, and spherical cavities can be treated analytically straightforwardly; realistic cavities are often variations on these shapes. We do not consider spherical cavities, as they typically do not have pairs of nearly degenerate modes.\footnote{It might be possible to use spherical cavities with the poles cut off, where the only modes that can be supported are nearly degenerate high harmonics. Alternatively,  one could couple two spherical cavities with a small tunable aperture as in Refs.~\cite{Ballantini:2005am, Bernard:2001kp, Bernard:2002ci}.} Furthermore, it is difficult to manufacture rectangular cavities with the required large quality factors. We hence focus on cylindrical cavities. 

An ordinary cylindrical cavity supports transverse electric ($\text{TE}_{mnp}$) and transverse magnetic ($\text{TM}_{mnp}$) modes, indexed by integers $m$, $n$, and $p$, as described in Appendix~\ref{sec:overlapints}. Because the axion carries no spin, and we have neglected its spatial gradients, it can only mediate transitions between modes with the same $m$. Furthermore, since the axion is a pseudoscalar, it must change the parity of $p$. Finally, axions cannot mediate transitions between pairs of TM modes. 

A simple option would be to use transitions between the two polarizations of a single TE mode, after splitting them in frequency by perturbing the cavity. However, this cannot work because the axion transition must change the parity of $p$. Instead, since the frequencies of the modes each depend differently on the cavity radius $R$ and length $L$, two modes could be arranged to be nearly degenerate by manufacturing the cavity with an appropriate aspect ratio $L/R$. As discussed further in Appendix~\ref{sec:overlapints}, overlap factors of $\eta_{10} \simeq 0.5$ can then be achieved for the transitions $\text{TM}_{0,n+1,0} \leftrightarrow \text{TE}_{0n1}$. For example, for a cavity loaded in the $\text{TM}_{030}$ mode, the loaded mode frequency is $\wl = 2 \pi \, \text{GHz}$ if the cylinder has radius $R \simeq \unit[0.4]{m}$. The $\text{TE}_{021}$ signal mode is degenerate if the length is $L \simeq \unit[0.25]{m}$, and a frequency difference of $m_a \sim \, \text{GHz}$ is attained if $L \simeq \unit[0.21]{m}$. Thus, many orders of magnitude in axion mass can be scanned by tuning the length through a relatively small range. 

Larger overlap factors of $\eta_{10} \simeq 0.8$ can be achieved by corrugating the outer wall with ridges. Similarly, using orthogonally oriented ridges on the end-walls of a square cross-section cavity to align the electric and magnetic fields of cross-polarized TE$_{10p}$/TE$_{01p}$ modes can also provide a large overlap factor, limited by how large/overmoded the cavity is. To further improve the quality factor, one can do the same with the cross-polarized HE$_{11p}$ hybrid modes in a cylindrical cavity with outer wall corrugations. This final approach is mathematically developed in Appendix~\ref{sec:app}. 

We now turn to physical mechanisms for tuning the frequency difference $\w_1 - \w_0$. Small changes in the cavity length can be achieved by applying pressure on the end-walls with a piezoelectric device. Concretely, the smallest scan steps we consider in Section~\ref{sec:reach} are of order $\unit[0.1]{Hz}$, which corresponds to changes in length of order $\unit[0.1]{nm}$. This tuning mechanism can deform a meter-long cavity by a few millimeters at most, leading to a scannable range of axion masses of about $\sim$ MHz. Larger changes in the cavity length can be achieved with mechanically retractable fins, as shown in Figure~\ref{fig:reach}(a). For non-corrugated cylindrical cavities, these fins effectively serve to change the length $L$ of the cavity, while for corrugated cylindrical cavities, they change the length seen by only one of the hybrid mode polarizations. 

Using fins, one can cover the full parameter space shown in Figure~\ref{fig:genreach} with a single cavity. However, introducing such sharp features into the cavity increases the peak surface fields, and hence has the potential to degrade the quality factor and lead to enhancement of field emission, as discussed in Section~\ref{sec:noiseDarkCurrent}. 

Since detailed numeric simulations of the cavity are required to understand these effects, we defer further discussion to future work. As such, the reach shown in Figure~\ref{fig:genreach} should be interpreted as indicating the potential of our general approach. However, we note that even an uncorrugated cylindrical cavity tuned solely with piezoelectric devices can probe a wide range of motivated parameter space, over orders of magnitude in axion mass. 

Finally, loading and readout can be achieved either through coaxial antennae fed into the cavity or with waveguides. For concreteness, we will employ the term ``waveguide" when discussing the loading/readout architecture. When we discuss the reach of the proposed approach in Section~\ref{sec:reach}, we will explore the optimization of the readout architecture. The language of waveguides lends itself well to this discussion, but the conclusions we reach do not depend on what specific instrument is used to extract the signal from the cavity. 

To summarize, as reference cavity parameters we consider $V\sim {\rm m}^3$ sized cylindrical SRF cavities operating at frequencies of $\omega/2\pi \sim {\rm GHz}$, with typical magnetic fields of $B \sim 0.2 \text{ T}$, and intrinsic quality factors of $Q_{\rm int} \gtrsim 10^{9}$. The level of frequency stability of modes planned for similar SRF cavities~\cite{fnalex} suggests that scanning step sizes of $\sim0.1 \text{ Hz} -1 \text{ Hz}$ are achievable. We therefore limit our analysis to frequency steps of $0.1$ Hz and above, and do not consider axion masses corresponding to frequencies below 1 Hz, where the effects of such a frequency instability become more dramatic. Furthermore, we do not consider the possibility of large frequency separations between the pump and signal mode, since this would involve accounting for intermediate modes. We therefore restrict our analysis to $m_a \lesssim \text{GHz}$.\footnote{An initial exploration of the use of higher harmonics of a loaded cavity was conducted in Ref.~\cite{Sikivie:2010fa}.}

\section{Noise Sources} \label{sec:noise}

\begin{figure}[t]
\includegraphics[scale=0.4]{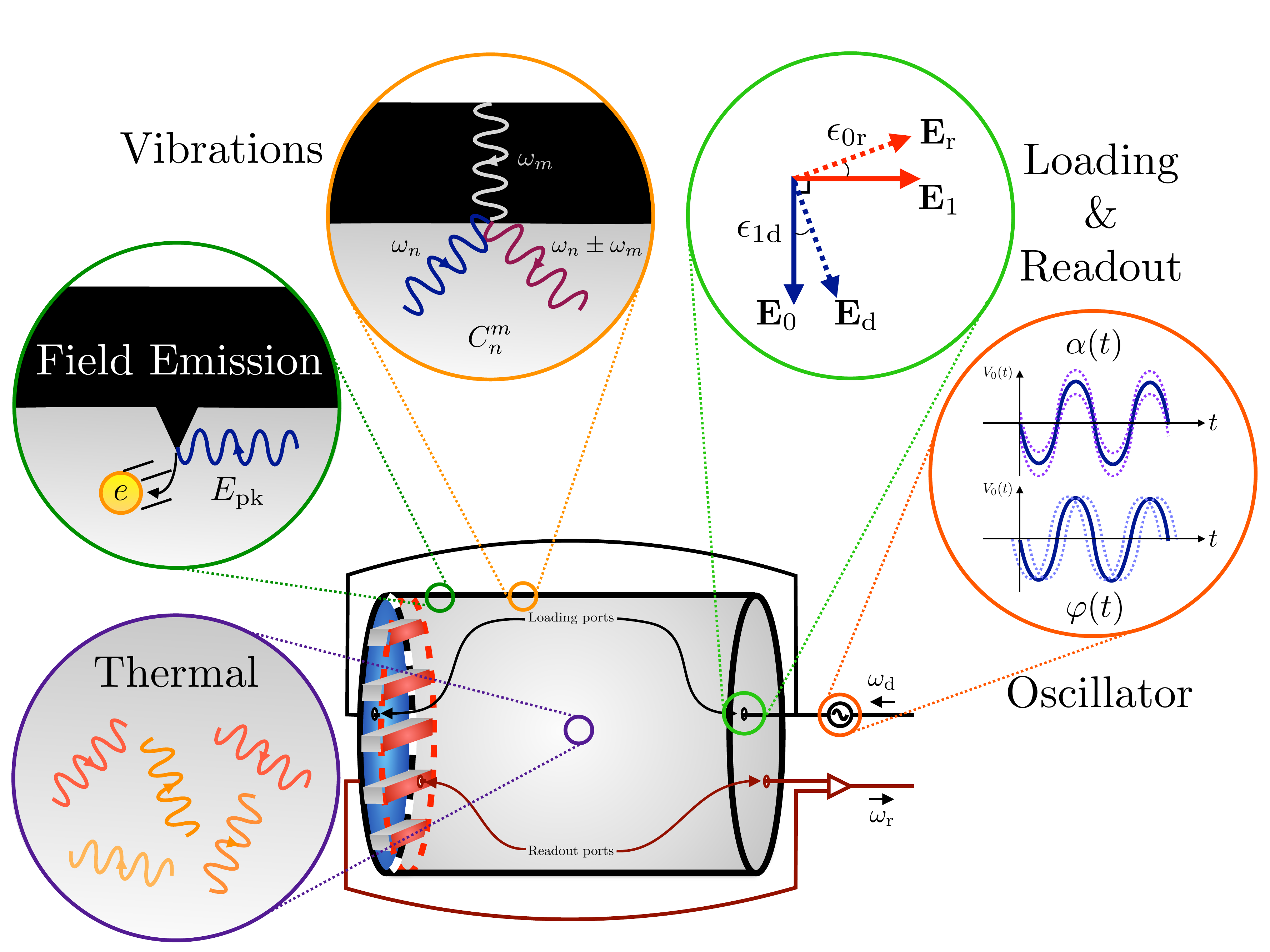}
\caption{
\label{fig:noisediagram}
A diagram depicting the main expected sources of noise specific to our detection strategy. In counterclockwise order are depictions of individual noise sources: thermal emission, discussed in Section~\ref{sec:noiseQuantum}; the effects of oscillator phase noise, as discussed in Section \ref{sec:leakage}; the precision of the geometric coupling of the loading and readout waveguides, relevant to several noise sources; vibrations of the cavity walls, discussed in Section~\ref{sec:noiseVibration}; and field emission, discussed in Section~\ref{sec:noiseDarkCurrent}. Not shown is amplifier noise, discussed in Section~\ref{sec:noiseQuantum}.}
\end{figure}

In this section, we describe the expected dominant noise sources for our setup, shown schematically in Figure~\ref{fig:noisediagram}. Some of these noise sources, such as amplifier and thermal noise, are common to axion DM experiments using static background magnetic fields~\cite{Boutan:2018uoc,Du:2018uak, Blout:2000uc, Brubaker:2016ktl,PhysRevLett.59.839,Wuensch:1989sa,Hagmann:1990tj, Zhong:2018rsr}. The remaining contributions, however, are particular to our setup. These include phase noise from the master oscillator that drives the pump mode, mechanical vibrations of the cavity walls, and field emission, commonly known in the accelerator community as ``dark current.'' 

\begin{figure}[t]
\centering
\subfloat[][$\epsilon_{1\text{d}} = 10^{-7}$, $Q_{\rm int} = 10^{12}$]{
\includegraphics[width = 0.48\textwidth]{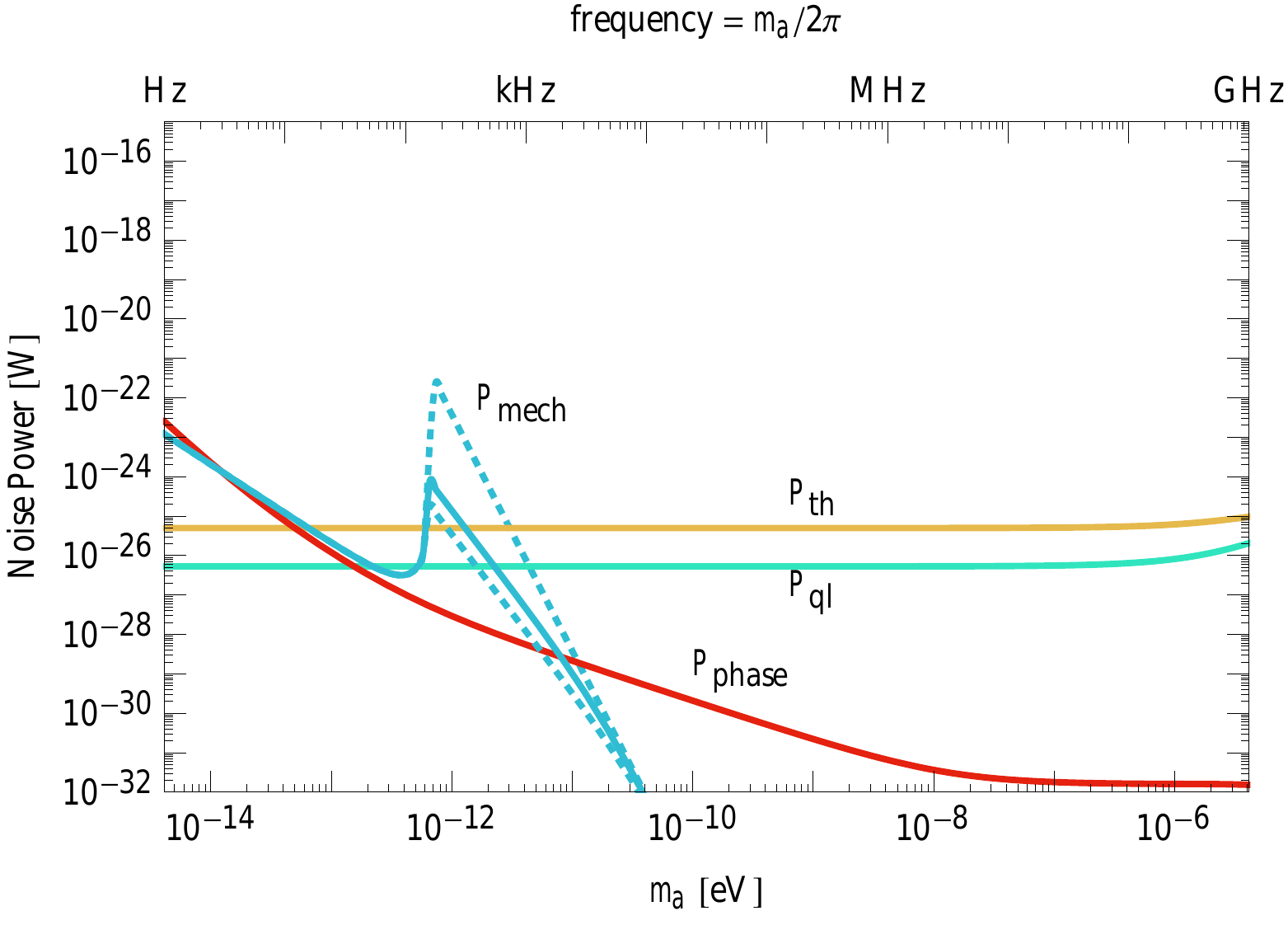}
}
\subfloat[][$\epsilon_{1\text{d}} = 10^{-5}$, $Q_{\rm int} = 10^{10}$]{
\includegraphics[width = 0.48\textwidth]{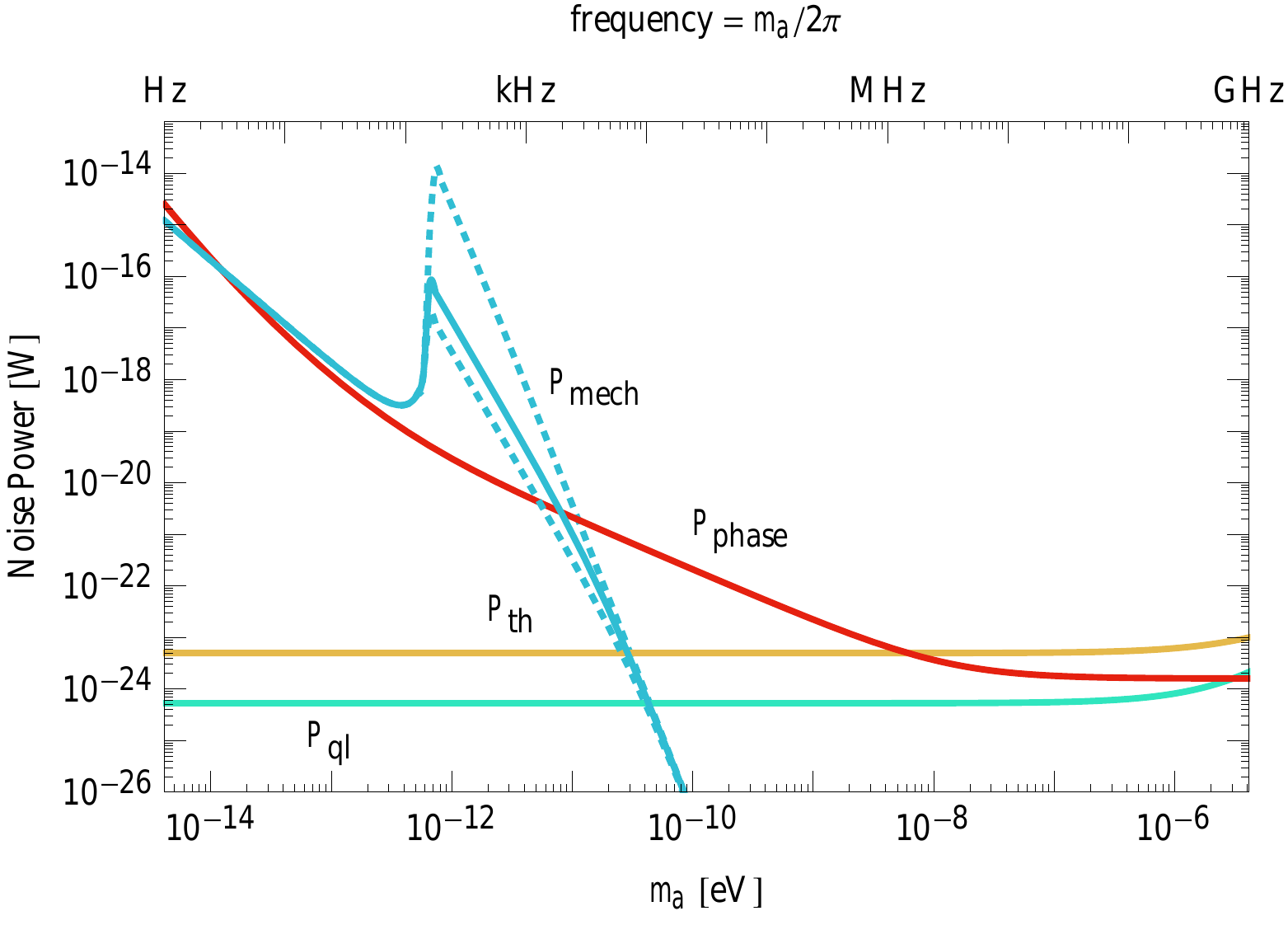}
}
\caption{
Comparisons of total power in thermal (yellow), amplifier (cyan), oscillator phase (red), and mechanical vibration (blue) noise, shown as a function of the axion mass $m_a$. The cavity parameters match the (a) lower and (b) upper curves in Figure~\ref{fig:genreach}. The figure shows the total power delivered to the readout architecture assuming critical coupling, and thus has appropriate factors of $Q_n / Q_{\rm cpl}$ included as discussed in Section~\ref{sec:reach}. The estimated size of mechanical noise depends on the degree of degeneracy between the axion mass and the resonant frequency of mechanical modes of the cavity. The solid line corresponds to the same model incorporated into the reach shown in Figures~\ref{fig:genreach} and \ref{fig:four_plot}, while the dashed lines serve to bracket the variation in such noise, depending on the scan/instrumental strategy employed (see Section~\ref{sec:noiseVibration} for discussion).
}
\label{fig:NoiseSources}
\end{figure}

The relative sizes of the noise sources, as a function of axion mass, are shown in Figure~\ref{fig:NoiseSources}. Thermal noise in the cavity, and amplifier noise in the readout system are both independent of $m_a$. Of the two sources, thermal noise in the cavity dominates, and plays the most important role at the largest axion masses that we consider. At smaller axion masses, two other sources of noise become relevant: frequency instability of the resonant modes from mechanical vibrations and power leakage from the pump to the signal mode. These both grow as the axion mass is decreased. As we discuss in the following, they are also both strongly sensitive to the quality factor of the cavity. Increasing the quality factor, other than increasing the signal power, decreases these two sources of noise. In  Figure~\ref{fig:NoiseSources}, the sharp feature evident in the mechanical noise power is due to our assumptions, motivated by the experimental characterization of similar cavities performed in Ref.~\cite{Bernard:2001kp}; we assume that there exists a spectrum of mechanical resonances above a kHz, each maximally coupled to the pump and signal modes of the cavity.

Before turning to a more detailed description of each of these noise sources, it is useful to distinguish the two contributions to the quality factor $Q_1$ of the signal mode,
\begin{equation}
\label{eq:Qsum}
\frac{1}{Q_1} = \frac{1}{Q_{\text{int}}} + \frac{1}{Q_{\text{cpl}}}
\, ,
\end{equation}
where $Q_{\text{int}}$ depends only on losses intrinsic to the cavity (such as the residual resistance of the walls) and $Q_{\text{cpl}}$ is determined by the rate at which power is transmitted to the readout. Critical coupling occurs when the two losses are equal, $Q_{\text{int}} = Q_{\text{cpl}}$, but we will see in Section~\ref{sec:reach} that it is optimal to strongly overcouple, $Q_1 \simeq Q_{\text{cpl}} \ll Q_{\text{int}}$, even though this degrades the total signal power. The readout is set to predominantly couple to the signal mode, as discussed further in Section~\ref{sec:leakage}, so that the pump mode's quality factor is not affected, $Q_0 \simeq Q_{\text{int}}$. The PSDs derived in this section represent the total noise power delivered to the cavity and to the readout apparatus in the signal mode. 

\subsection{Thermal and Amplifier Noise} \label{sec:noiseQuantum} \label{sec:noiseThermal}
%!TEX root = axion_arxiv.tex

Thermal emission of radio waves from the the cavity walls constitutes an irreducible noise source. If the cavity is cooled to a temperature $T$, then the PSD of this thermal noise is
\be
\label{eq:thermal}
S_\text{th} (\w) = \frac{Q_1}{Q_{\text{int}}} \, \frac{4\pi T \, (\w \, \w_1 / Q_1)^2}{(\w^2 - \w_1^2)^2 + (\w \, \w_1 / Q_1)^2} 
~.
\ee
Here, the prefactor of $4\pi$ stems from our use of two-sided PSDs and the convention of Eq.~\eqref{eq:PSDdef}, and the factor of $Q_1 / Q_{\text{int}}$ arises because the coupling to the readout does not source thermal noise; it is only the cavity walls that are at temperature $T$. This corresponds to an average total noise power of
\be
P_\text{th} \simeq \frac{T \, \w_1}{Q_{\text{int}}}
\, .
\label{eq:Pthermal}
\ee
Driven SRF cavities can be efficiently cooled using a superfluid helium bath to a temperature of $T = \unit[1.8]{K}$. This temperature is below the superfluid transition at $\unit[2.2]{K}$, which mitigates vibrational noise from the bubbling of gaseous helium. Dissipation of the pump mode increases the temperature of the cavity walls slightly above that of the helium bath, but we neglect this since the typical temperature change is small, of order $\unit[0.1]{K}$~\cite{padamsee2009rf}.

The signal is to be read out with an amplifier coupled to the resonant cavity. We assume that amplifier noise can be reduced to its standard quantum limit, resulting in one photon of power per unit bandwidth~\cite{PhysRevB.70.245306, PhysRevD.26.1817}. One half of this power is due to quantum zero-point fluctuations, while the other half accounts for the backaction and imprecision noise associated with the amplifier~\cite{RevModPhys.82.1155}. The corresponding noise power can be described as a spectrally flat PSD of the form
\be
S_{\rm ql} (\w) = 4\pi \w_1 
\, .
\ee
This assumption is equivalent to that made for other experiments targeting similar axion parameter space~\cite{Chaudhuri:2019ntz} and has been achieved in practice at GHz frequencies~\cite{Brubaker:2016ktl}. For context, at critical coupling, amplifier noise is smaller than thermal noise by a factor of the thermal occupation number, $n_{\text{occ}} = T / \w_1 \sim 100$. 

Overcoupling the cavity to the readout can enhance the reach of a thermal noise limited search~\cite{Chaudhuri:2019ntz}. Intuitively, this is possible since a quantum-limited readout has an effective noise temperature given by a single photon of noise per unit bandwidth ($T_{\text{eff}} \sim \omega_1 \sim 10 \text{ mK} \ll \unit[1.8]{K}$), and so overcoupling lowers the effective noise temperature of the system. Similar statements can be made when other noise sources dominate. We discuss these aspects in more detail in Section~\ref{sec:reach} and Appendix~\ref{sec:overcouple}.

\subsection{Oscillator Phase Noise} \label{sec:leakage}
%!TEX root = axion_arxiv.tex

The pump mode is excited by driving a waveguide at frequency $\w_0$ with an external oscillator. The loading waveguide possesses a geometric coupling to the pump mode $\epsilon_{0\text{d}} \simeq 1$ and is adjusted to have a small coupling to the signal mode, $\epsilon_{1\text{d}} \ll 1$. Similarly, the signal is detected through a readout waveguide, which is adjusted to have a small coupling to the pump mode, $\epsilon_{0\text{r}} \ll 1$. Since minimizing the unwanted geometric couplings $\eps_{1 \text{d}}$ and $\eps_{0 \text{r}}$ requires precisely controlling the geometry of the two waveguides, we take $\eps_{1 \text{d}} \simeq \eps_{0 \text{r}}$. The  mechanical precision required to achieve a certain rejection value is discussed further in Appendix~\ref{app:rejection}.

The oscillator is centered around the frequency $\omega_0$, but is broadened due to fluctuations in the amplitude and phase of its output voltage, which can be parametrized as 
\be
V_\text{osc}(t) = V_0 \, (1 + \alpha(t)) \cos(\w_0 t +\varphi(t))
\,.
\ee
The PSD of the amplitude noise $S_\alpha(\w)$ typically has flat (white) and $1/\w$ components, the latter due to so-called ``flicker noise.'' The PSD of the phase noise $S_\varphi(\w)$
has additional $1/\w^2$ and $1/\w^3$ components due to the Leeson effect, whose effects dominate over amplitude noise for the small frequency splittings that we consider~\cite{rubiola2009phase,Pozar:882338}. The component of $V_{\text{osc}}(t)$ at frequency $\omega_1$ can be inadvertently read out as signal through the coupling $\epsilon_{1\text{d}}$ or $\epsilon_{0\text{r}}$. 

Oscillator manufacturers typically report the one-sideband noise power per unit bandwidth, relative to the carrier power. From this we extract the phase noise PSD $S_\p(\w)$.
We fit the reported spectrum of a low-noise commercially available oscillator~\cite{datasheet} to the functional form 
\be
S_\p(\w) = \sum_{n=0}^3 b_n \, \w^{-n}
\, ,
\ee
yielding the values 
\be
b_0 = 10^{-16} \Hz^{-1}
~,~
b_1 = 10^{-9}
~,~
b_2 = 10^{-6} \Hz
~,~
b_3 = 10^{-5} \Hz^2
~.
\ee
Defining the total power input to the cavity as
\begin{equation}
P_{\text{in}} = \frac{\omega_0}{Q_0} \, B_0^2 V
\, ,
\end{equation}
the PSD due to oscillator phase noise is given by 
\be
S_{\text{phase}}(\w) \simeq \frac12 \, \eps_{1 \text{d}}^2 \, S_\p (\w - \w_0) \, \frac{(\w \, \w_1 / Q_1)^2}{(\w^2 - \w_1^2)^2 + (\w \, \w_1 / Q_1)^2} \, \frac{\wl Q_1}{\ws Q_0} \, P_{\text{in}} 
\, .
\label{eq:Sphase}
\ee
Note that because the width $\omega_0 / Q_0$ of the pump mode is much smaller than the axion mass $m_a = \omega_1 - \omega_0$ for all parameters we consider, the noise due to the coupling $\epsilon_{0\text{r}}$ is suppressed by the Breit--Wigner tail of the pump mode, $(\omega_0 / m_a Q_0)^2 \ll 1$, and is hence negligible. Taking $S_\p(\w)$ to be spectrally flat within the signal mode bandwidth, which is a good approximation for all parameters shown in Figure~\ref{fig:genreach}, the above PSD corresponds to an average total noise power of
\be
P_\text{phase} \simeq \frac{\eps_{1 \text{d}}^2\, S_\p(m_a)}{16 \pi} \, \frac{\w_0}{Q_0} \, P_{\text{in}}
\, .
\label{eq:Pphase}
\ee

Projected sensitivities are shown in Figure~\ref{fig:genreach} for $\eps_{1\text{d}}=10^{-5}$, $10^{-7}$. Geometric rejections at the level of $\order{10^{-7}}$ have been experimentally demonstrated in Refs.~\cite{Bernard:2001kp, Ballantini:2005am} for a different signal and pump mode geometry. As discussed in greater detail in Appendix~\ref{app:rejection}, achieving $\epsilon_{1\text{d}}=10^{-7}$ in our setup requires controlling the cavity components at the few nm level, which is a level of precision that is already envisioned for other applications~\cite{fnalex}. For such small rejection factors, $\epsilon_{1\text{d}}$ scales linearly with this distance scale. As shown in Figure~\ref{fig:NoiseSources}, we find that phase noise is subdominant compared to thermal noise for the largest axion masses that we consider, while it dominates at smaller masses.

\subsection{Mechanical Vibration Noise} \label{sec:noiseVibration}
%!TEX root = axion_arxiv.tex

Mechanical oscillations of the cavity boundaries lead to time-dependent shifts in the resonant modes and their corresponding frequencies. Such perturbations can impede the ability to reliably scan over the axion mass range and may also induce transitions between the pump and signal modes, thus constituting a potential background to the axion-induced signal. Various forces can contribute to mechanical noise such as thermal excitations of the cavity, external vibrations from the cryogenic cooling system or seismic activity, and radiation pressure due to the electromagnetic energy stored in the loaded mode. Of these sources, the last is negligible, because it does not source significant vibrations at frequency $\order{m_a}$. Instead, its dominant effect is to introduce a static shift in the cavity mode frequencies, known in the accelerator community as ``Lorentz force detuning," which we may simply absorb into the definitions of $\w_0$ and $\w_1$. Thermal effects are irreducible but, as we will argue below, subdominant, while  power from external sources can be significantly attenuated through active feedback or isolation of the suspended cavity from its immediate surroundings.
 
To estimate both thermal and vibrational effects, we follow the discussion in Ref.~\cite{Bernard:2002ci}. The displacement of the cavity wall from its equilibrium position, denoted as $\uv(\xv, t)$, can be decomposed as a sum over the various dimensionless mechanical normal modes of the cavity, $\xiv_\alpha (\xv)$,
\be
\label{eq:mechnorm}
\uv(\xv, t) = q_\alpha(t) \, \xiv_\alpha (\xv)
~,
\ee
where the expansion coefficients are given by the time-dependent generalized coordinates, $q_\alpha(t)$, and a sum over the integer $\alpha$ is implied. The mode vectors are normalized such that
\be
\int d^3 \xv ~ \rho (\xv) ~ (\xiv_\alpha \cdot \xiv_\beta) = M \, \delta_{\alpha \beta}
~,
\ee
where $\rho$ and $M$ are the mass density and total mass of the cavity, respectively.  In the following, we focus on an individual mechanical resonance, labeled by $\alpha = m$.  The noise power from multiple resonances can be summed, but in most cases only the nearest resonance will be relevant.  The response of the generalized coordinate of the cavity boundary is described by the PSD,
\be
\label{eq:qmPSD}
S_{q_m} (\w) \simeq \frac{1}{M^2} ~ \frac{S_{f_m} (\w)}{(\w^2 - \w_m^2)^2 + (\w_m \w / Q_m)^2}
~, 
\ee
where $\w_m$ is the resonant frequency of the excited mechanical mode, $Q_m$ is the mechanical quality factor, and $f_m$ is the force projected onto mode $\alpha=m$. In our estimates, we adopt $Q_m = 10^3$ as a representative value~\cite{Ballantini:2005am}. The force term on the right-hand side of Eq.~(\ref{eq:qmPSD}) contains contributions from radiation pressure, thermal fluctuations, and other environmental sources.

The force generated from thermal fluctuations is negligible compared to seismic or cryogenic noise for realistic attenuation capabilities. For the cavity parameters we consider, thermal vibrations source 
$S_{f_m^\text{th}} (\w) \sim 10^{-23} \ \text{N}^2 / \text{Hz} \times \left(M/\text{kg}\right) \left(T/\text{K} \right)  \left(\w_m/\text{kHz} \right) \left(10^3/Q_m\right)$, while, e.g., the authors of Ref.~\cite{Ballantini:2005am} directly measured the \emph{unattenuated} force PSD for a similar resonant cavity design and found values spanning from $\order{10^{-7}} \text{ N}^2 / \text{Hz} - \order{10^{-3}} \text{ N}^2 / \text{Hz}$ within the measured frequency range of  $10 \text{ Hz} - 10 \text{ kHz}$, stemming from vibrations of the surrounding environment. For realistic attenuation factors, the latter vibrational sources are dominant. 

Rather than directly reporting an attenuated external force PSD, experiments frequently characterize mechanical noise by the RMS wall displacement $q_\text{rms}$ induced by these forces.  For example, near-term light-shining-through-wall type experiments at FNAL plan on controlling wall displacements of loaded cavities to within sub-nanometer precision through the use of piezo-actuator tuners~\cite{fnalex}.  To infer a force PSD from this level of vibration, we note that \eq{eq:qmPSD} implies an RMS displacement of the $m$'th normal mode %
\be
\label{eq:wobble}
 \langle q_m^2 \rangle \simeq \frac{S_{f_m} (\w_m) \, Q_m }{4 \pi \, M^2 \, \w_m^3} \sim 10^6 \text{ nm}^2 \times \left( \frac{S_{f_m}(\w_m)}{10^{-4} \ \text{N}^2 \ \text{Hz}^{-1}} \right) \left( \frac{Q_m}{10^3} \right) \left( \frac{M}{\text{kg}} \right)^{-2} \left( \frac{\w_m}{\text{kHz}} \right)^{-3}
~.
\ee
The scaling with $\w_m$ implies that, for an approximately flat $S_{f_m}(\w)$,  $\langle q_m^2 \rangle$ is largest for the lowest-frequency mechanical mode.  Thus, we normalize the attenuated force PSD to
\be
S_{f_m} \simeq 4 \pi \, M^2 \, \w_{\text{min}}^3 \, q_\text{rms}^2  / Q_m
~,
\ee
where $q_\text{rms} \sim 0.1 \text{ nm}$ and $\w_{\text{min}} \sim \text{kHz}$ is the lowest-lying mechanical resonance of the cavity. This estimate of $\w_\text{min}$ is motivated by the measurements of a similar apparatus to search for gravitational waves, which showed a growing number of mechanical resonances above $\w \sim 0.5 \text{ kHz}$~\cite{Bernard:2001kp}. Hence, we will assume that externally sourced vibrations are controlled to $\langle q_m^2 \rangle \sim (0.1 \text{ nm})^2$, which from \eq{eq:wobble} implies an attenuation ability of $\order{10^{-8}}$. Note that even assuming a considerably worse control of the cavity walls, $q_\text{rms} \sim 10^2\;\mu{\rm m}$, the estimated sensitivity at large axion masses, and in particular the ability to probe the QCD axion, is not appreciably affected, as shown in Figure~\ref{fig:four_plot}.

These mechanical vibrations couple to the electromagnetic cavity modes by, e.g., shifting their resonant frequencies,
\be
\label{eq:freqresc}
\delta \w_n (t) \simeq - \frac{1}{2} \, q_m (t) \, C_n^m \, \w_n 
~,
\ee
where the coupling coefficients, $C_{n}^m$, are given in terms of the electromagnetic modes,\footnote{We have assumed that the off-diagonal generalizations of the coupling coefficient involving pairs of distinct electromagnetic modes vanish to leading order in the cavity perturbation. We have checked that this is satisfied for various nearly-degenerate modes of cylindrical cavities, which have orthogonal ${\mathbf{E}}$ and ${\mathbf{B}}$ fields at every point in space. If this is not the case, additional source terms in the coupled electromagnetic-mechanical equations of motion should be included. See Ref.~\cite{Bernard:2002ci} for additional details.}
\be
\label{eq:Cexp}
C_{n}^m = \frac{\int d \Sv \cdot \xiv_m(\xv) ~ \big( |\mathbf{B}_n (\xv)|^2 - |\mathbf{E}_n (\xv)|^2 \big)}{\int d^3 \xv ~ |\mathbf{E}_n(\xv)|^2}
~.
\ee
In the numerator of \eq{eq:Cexp}, the integral is performed over the surface boundary of the deformed cavity. Note that the size of the coupling coefficient $C_n^m$, and hence also the frequency shift of \eq{eq:freqresc}, depends on the specific nature of the mechanical and electromagnetic resonances of the unperturbed cavity. We will pessimistically assume maximum overlap between the mechanical and electromagnetic modes, in which case the coupling coefficient is parametrically of size $C_{n}^m \sim V^{-1/3}$, where $V$ is the geometric volume of the cavity. 

The shift in the cavity mode frequencies in \eq{eq:freqresc} results in a modification of the equation of motion for a mode (labeled $n$) driven by an external field $\Dvec(t, \mathbf{r})$, 
\be
\Big[\partial_t^2 + \frac{\w_n}{Q_n} \, \partial_t + (\w_n + \delta \w_n)^2  \Big] \, \Bvec_n(t, \mathbf{r}) = \w_n^2 \, \Dvec(t, \mathbf{r}) \ .
\label{eq:VibEoM}
\ee
When the time-dependent shifts in the cavity mode frequencies are small, we can perturbatively solve the above equation to find the noise PSD due to vibrations of the cavity walls, $S_\text{mech} (\w)$. 
To leading order in $\delta \w_n^2 / \w_n^2 \ll 1$, we find
\be
S_\text{mech} (\w)= \sum_{n=0,1}S_\text{mech}^{(n)} (\w)\simeq \frac{\eps_{1 d}^2}{4} \, \frac{\w_0}{Q_0} \, P_{\text{in}} \sum\limits_{n=0,1} \frac{\left(S_{q_m}(\w - \w_0)/V^{2/3}\right)(\w_n / Q_n) \, \w_n^4 \, \w^2}{\big[ (\w^2 - \w_n^2)^2 + (\w \, \w_n / Q_n)^2\big] \, \big[ (\w_0^2 - \w_n^2)^2 + (\w_0 \, \w_n / Q_n)^2\big] }\, , 
\label{eq:MechPSD}
\ee
where the sum is over the pump ($n=0$) and signal ($n=1$) modes. 
To understand Eq.~\eqref{eq:MechPSD} parametrically, we note that for $m_a \simeq \w_m$ and $\w_m / Q_m \ll \w_n / Q_n$, evaluating $S_\text{mech}(\w)$ near the positive frequency resonance ($\w \simeq \w_1$) and applying Eqs.~(\ref{eq:qmPSD}) and (\ref{eq:wobble}) yields
\be
S_\text{mech} (\w_1+\Delta \w) \simeq \frac{\pi}{2} \, \frac{\eps_{1d}^2 \, Q_m}{1 + ( \Delta \w / \Delta \w_m)^2} \, \frac{\w_1^2 \w_{\text{min}}^3}{m_a^6} \, \frac{q_\text{rms}^2}{V^{2/3}} \, P_\text{in}
~,
\ee
where we defined the width of the mechanical mode $\Delta \w_m \equiv \w_m / 2 Q_m$.  

In the coupled superconducting cavity setup of Ref.~\cite{Bernard:2001kp}, direct probes of the designed apparatus revealed the presence of mechanical resonances above $\w_\text{min} \sim \text{kHz}$, separated in frequency by $\order{100} \text{ Hz}$. For $m_a < \w_\text{min}$, mechanical noise is driven by the tail of the lowest-frequency resonance. In this regime, the scaling of mechanical noise is dominated by the cavity's response to an off-resonance driving force, as expected from the form of Eqs.~(\ref{eq:qmPSD}) and (\ref{eq:MechPSD}). Therefore, the noise power scales as roughly $1/m_a^2$. 
 As shown in Figure~\ref{fig:NoiseSources}, mechanical noise is significant in this mass range and is roughly comparable to oscillator phase noise. 
  
The behavior of mechanical noise in the vicinity of resonances is more subtle, and so merits further discussion. 
Eqs.~(\ref{eq:qmPSD}) and (\ref{eq:MechPSD}) imply that the power in mechanical noise is maximized for $\w_m \simeq \w_1 - \w_0$. Thus, in a scan over $\w_1$, the mechanical noise PSD has a forest of local maxima around each resonance $\w_0 + \w_m$, with minima in between.  In Figure~\ref{fig:noisediagram}, we bound the total power in mechanical noise for $m_a > \w_\text{min}$ by considering two cases: where the axion mass is situated at or near a local maximum of the noise PSD ($m_a \simeq \w_m$ for some mechanical resonance $m$), or at a local minimum ($m_a$ at the midpoint between two adjacent resonances, i.e., assuming a typical separation of $\sim 100$ Hz between mechanical resonances, at 50 Hz separation from each).  The total mechanical noise powers obtained in these two extreme cases, illustrated by dashed curves in Figure~\ref{fig:noisediagram}, define an envelope for the mechanical noise power at each scan step.  
The envelope spans 3 orders of magnitude in noise power, due to the sharpness of the mechanical resonances, but for the same reason, the noise power only approaches the upper envelope in narrow regions of size $\Delta \w_m$ about each resonance.  

For a more representative characterization of the mechanical noise near resonances, we note that in a scan over the range of candidate axion masses between any two resonances, the median noise PSD is that obtained at 25 Hz separation from the nearest mechanical resonance.  The total mechanical noise power at this separation is indicated by the solid blue curve in Figure~\ref{fig:noisediagram}, and this characteristic noise power is used in deriving the axion sensitivity curves.  In a single scan, half of candidate axion masses are expected to have noise above this line (and hence weaker sensitivity) and half below (and hence stronger sensitivity).  It may also be possible to fill these narrow gaps in sensitivity by using two cavities with slight mechanical variations, so that their mechanical resonance frequencies are slightly offset.  In this case, each candidate axion mass will be well-separated from the mechanical resonances of at least one of the two cavities.

Near-resonance mechanical noise is only expected to dominate over about one decade in axion mass near angular frequencies of $1 \text{ kHz}-10 \text{ kHz}$.  At lower frequencies there are no nearby resonances, and mechanical noise falls off rapidly at higher frequencies.  These two effects lead to a peak-like structure near  $m_a = 1$ kHz in Figure~\ref{fig:NoiseSources}, and corresponding dips in the reach shown in Figures~\ref{fig:genreach} and \ref{fig:four_plot}.  The strength and position of this feature should be appropriately rescaled by $\w_\text{min}$ for cavities with higher- or lower-lying resonances.

For $m_a \ll \text{MHz}$, where mechanical noise is important, the integral of \eq{eq:MechPSD} over the signal bandwidth is analytically tractable for the pessimistic case of a mechanical resonance very closely spaced to $\w_1 - \w_0 \simeq m_a$. Taking the mechanical resonance to be narrower than the cavity bandwidth and further approximating $Q_0 \simeq Q_1$ and $m_a \ll \w_0$, the average total noise power in mechanical noise is
\be
P_\text{mech} \simeq \frac{\eps_{1d}^2}{16} \, \frac{\w_0^2 \, \w_{\text{min}}^3}{m_a^5} \, \frac{q_\text{rms}^2}{V^{2/3}} \, P_{\text{in}} 
\, .
\label{eq:Pmech}
\ee

We emphasize that the mechanical noise estimates presented above are most likely overly pessimistic. In particular, we assumed that for every axion mass $\gtrsim \text{kHz}$ there is a corresponding resonant mechanical mode that is maximally coupled to the electromagnetic properties of the cavity. In this sense, a dedicated design strategy could potentially significantly mitigate noise from mechanical vibrations.

\subsection{Field Emission} \label{sec:noiseDarkCurrent}
%!TEX root = axion_arxiv.tex

At high surface electric fields, electrons are emitted from imperfections on the walls of the SRF cavity. The released electrons accelerate to relativistic speeds, absorbing energy from the cavity field, and typically are reabsorbed into the wall within less than one oscillation cycle of the cavity. They emit radiation in three different stages: as they accelerate inside the cavity and emit synchrotron radiation; as they encounter the dielectric mismatch between the interior and wall of the cavity, leading to transition radiation; and as they encounter the nuclear electric fields of the wall material, leading to Bremsstrahlung radiation. 

In this section, we crudely approximate the noise due to each process. First, we note that for an electron of energy $\gamma m_e$, all three processes produce radiation in a small solid angle $1/\gamma^2$ around the electron momentum, spread roughly uniformly over a frequency range much broader than the signal mode bandwidth. Hence, only a small fraction of the power absorbed by the emitted electrons is deposited in the signal mode. 

The energy absorbed by a single electron as it traverses a cavity of length $L \sim 1/\w_0\sim \unit[1]{m}$ and average electric field of strength $E_0 \sim c B_0 = \unit[60]{MV/m}$ is roughly $U_{\text{abs}} \sim e E_0 L$, corresponding to a Lorentz factor of $\gamma \sim 100$. This energy is then released through the three processes described above. The energy released as synchrotron radiation and the corresponding frequency range are
\be
U_{\text{sync}} \sim \frac{e^4 \gamma^2 E_0^2 L}{\me^2} \, , \quad \Delta \w_{\rm sync} \gtrsim \w_0
\, ,
\ee
where the frequency range is determined by the short timescale $t_{\rm sync} \sim 1/\w_0$ over which the electron is within the cavity. The energy released in transition radiation depends on the plasma frequency of the wall material, which for niobium is $\w_p \sim \unit[50]{eV}$. The spectrum of the produced radiation is relatively flat, and the total energy released and the corresponding frequency range are~\cite{jackson_classical_1999} 
\be
U_{\text{trans}}\sim e^2 \gamma \w_p\, , \quad \Delta \w_{\rm trans} \sim \gamma \w_p
\, .
\ee
Finally, as the electron travels inside the cavity wall, the remaining energy is released through losses inside the material. Since both $U_{\text{sync}}$ and $U_{\text{trans}}$ are both much smaller than $U_{\text{abs}}$, almost all the absorbed energy is released inside the wall. We assume that all of the energy is converted into photons via Bremsstrahlung, that all of these photons are released into the body of the cavity, and that the spectrum of the radiation is approximately flat, giving
\be
U_{\text{brem}} \sim e E_0 L\, , \quad \Delta \w_{\text{brem}} \sim \gamma \me\, .
\ee
In reality, the release of energy in the walls is a complex process, which our assumptions model only very crudely. Our first two assumptions are very pessimistic, while our third assumption is optimistic, as a relativistic electron will create showers of softer electrons which release energy within a smaller frequency range. However, in any case, we will find that the noise PSD due to Bremsstrahlung is subdominant by several orders of magnitude. 

We can use these results to evaluate the noise PSD, normalized to the total power loss $P_{\text{tot}}$ due to field emission. For concreteness, we compare the three contributions to the typical PSD for thermal noise. Accounting for the small geometric overlap factor $1/\gamma^2$, the PSDs are
\be
S_i(\w) \sim P_{\text{tot}} \, \frac{U_i}{U_{\text{abs}}} \, \frac{1}{\gamma^2 \Delta \w_i} 
\, ,
\ee
where $P_{\text{tot}}$ is the total power loss due to field emission. Numerically, we have
\be
\label{eq:fieldPSD}
\frac{S(\w_1)}{4\pi T} \sim \frac{P_{\text{tot}}}{\unit[0.1]{W}} \times \begin{cases} 1 & \text{synchrotron} \\ 10^{-6} & \text{transition} \\ 10^{-5} & \text{Bremsstrahlung} \, , \end{cases}
\ee
so that for field emission to be negligible compared to thermal noise, we require $P_{\text{tot}} \lesssim \unit[0.1]{W}$. For context, this corresponds to $\order{100}$ electrons emitted per cycle, or about 0.1\% of the total energy loss for a cavity with $Q_{\text{int}} = 10^{12}$. 

In practice, the rate of field emission is set by the shapes of each cavity's particular defects, which determine the local enhancement of the electric field. Since it is a tunneling effect, the electric current due to a given defect has a strong exponential dependence on the field, $I \sim \text{exp}(-1/\beta E)$, where $\beta$ depends on the geometry of the defect~\cite{padamsee1998rf}. As such, field emission from a defect is essentially zero for lower fields, then sharply increases at a certain threshold field value. Modern cavity fabrication techniques can produce cavities where field emission is a small source of energy loss (defined as $P_{\text{tot}} < \unit[10]{W}$) up to peak surface electric fields beyond $\sim \unit[60]{MV/m}$. Moreover, in many cases, field emission is not even detectable for peak surface fields of this magnitude~\cite{wiener2008improvements,padamsee2009rf}.

Given this background, the relevance of field emission to our setup depends sensitively on the design. For the cylindrical cavity modes discussed in Section~\ref{sec:realisticCavity}, the peak surface electric fields are several times smaller than the typical fields $E_0 \sim \unit[60]{MV/m}$, making field emission a completely negligible effect. However, the use of retractable fins to tune the frequency difference would create a sharp feature within the cavity and hence a local enhancement of the surface field. As discussed in Section~\ref{sec:realisticCavity}, we are sensitive to a wide range of motivated parameter space even without the implementation of fins; we defer further discussion of field emission in this setting to a future detailed study of the experimental design. In particular, both field emission and the resulting synchrotron radiation can be simulated more precisely using existing  dedicated numeric programs. 

Finally, it is worth mentioning other well-known effects associated with SRF cavities. The cavity must be designed and manufactured to manage well-understood problems such as multipacting and thermal breakdown~\cite{padamsee1998rf,padamsee2009rf}. Another physical effect to consider is nonlinearity in the response of the cavity walls to the pump mode fields, which could produce radiation with frequency at integer multiples of $\w_0$. However, this is not relevant for our setup because the signal mode frequency $\w_1$ is not close to any of these multiples; instead we have $\w_1 \simeq \w_0$.

\section{Physics Reach}\label{sec:reach}
%!TEX root = axion_arxiv.tex

\begin{figure}[t]
\centering
\includegraphics[scale=0.64]{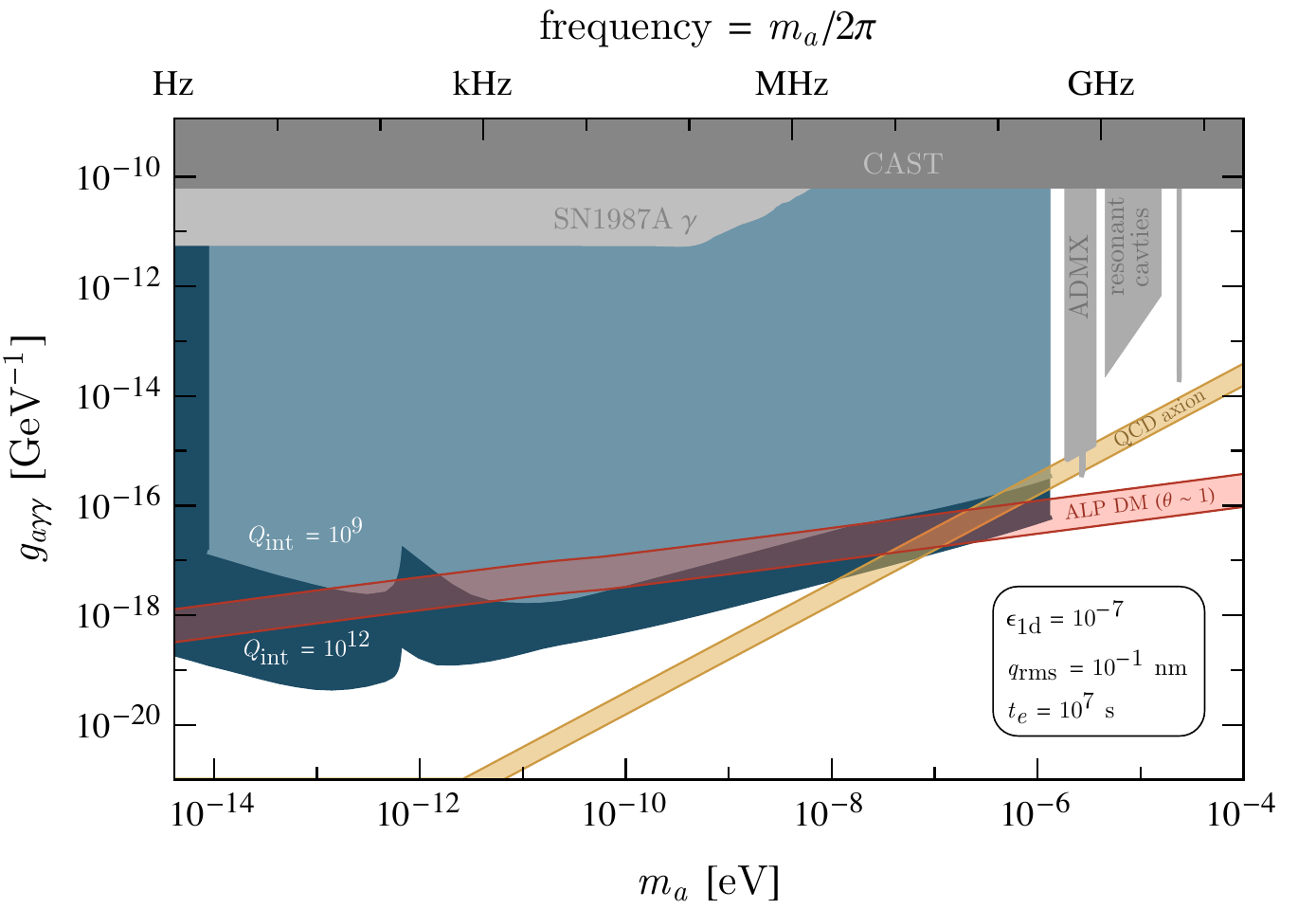}
\includegraphics[scale=0.64]{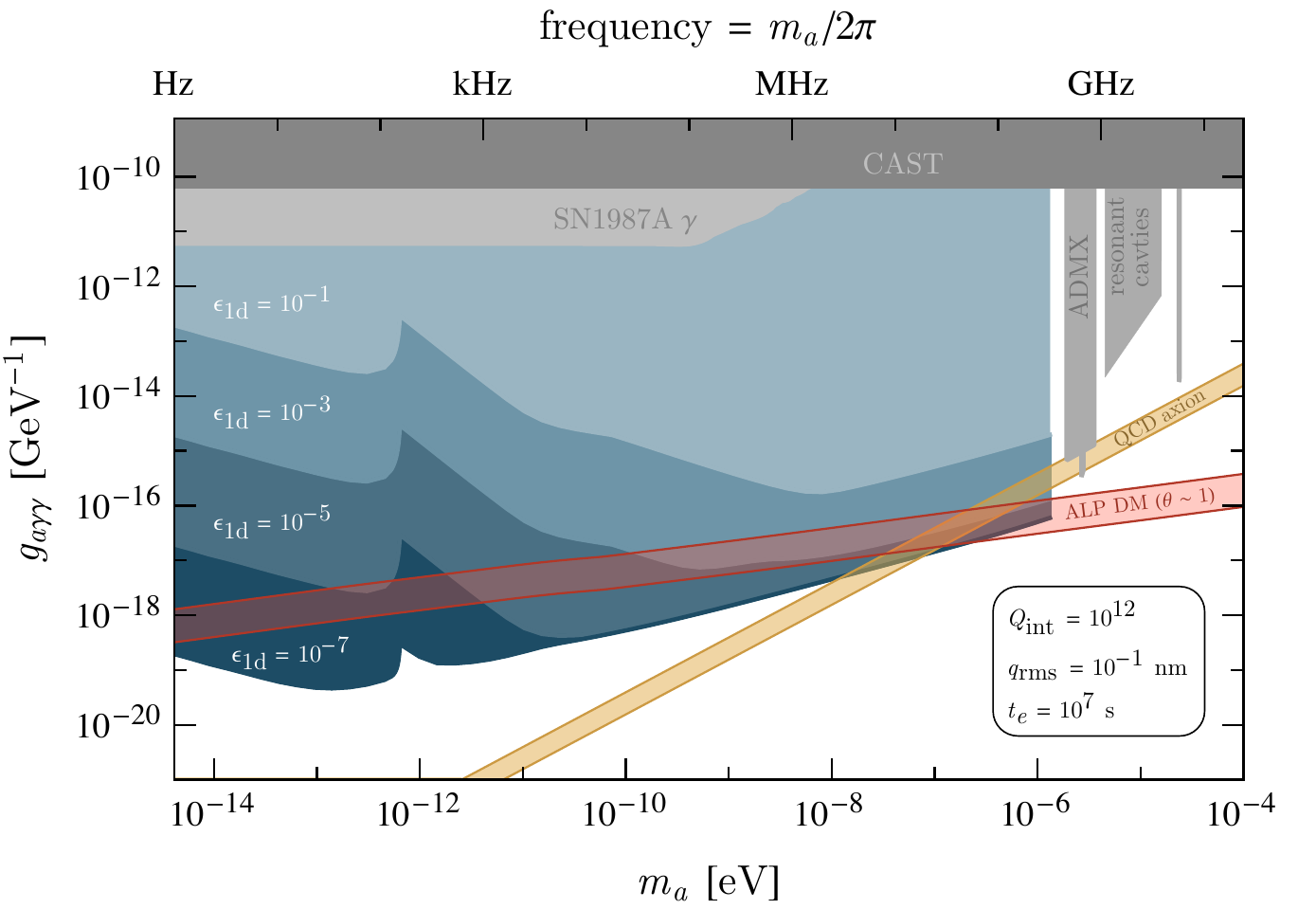}
\includegraphics[scale=0.64]{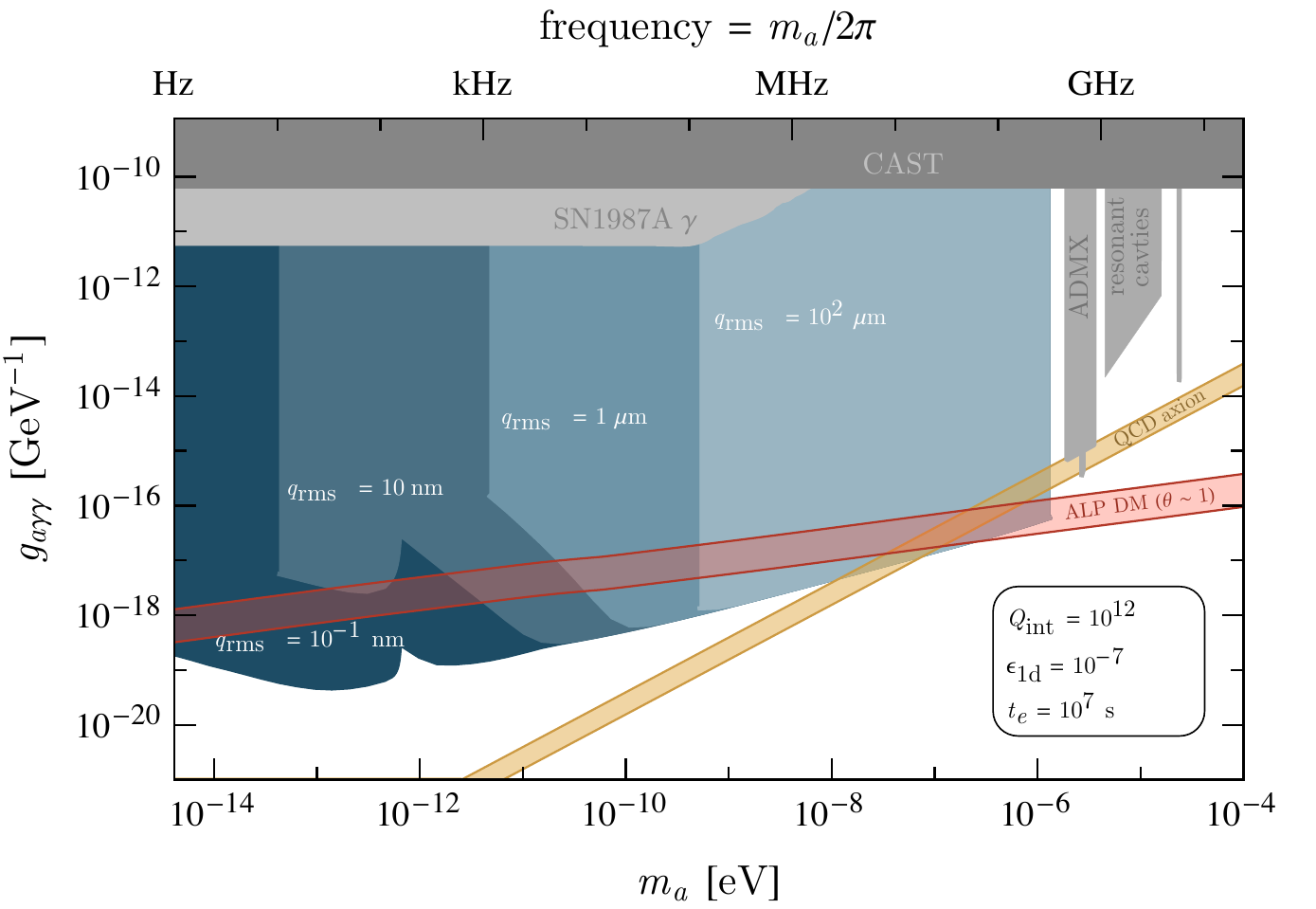}
\includegraphics[scale=0.64]{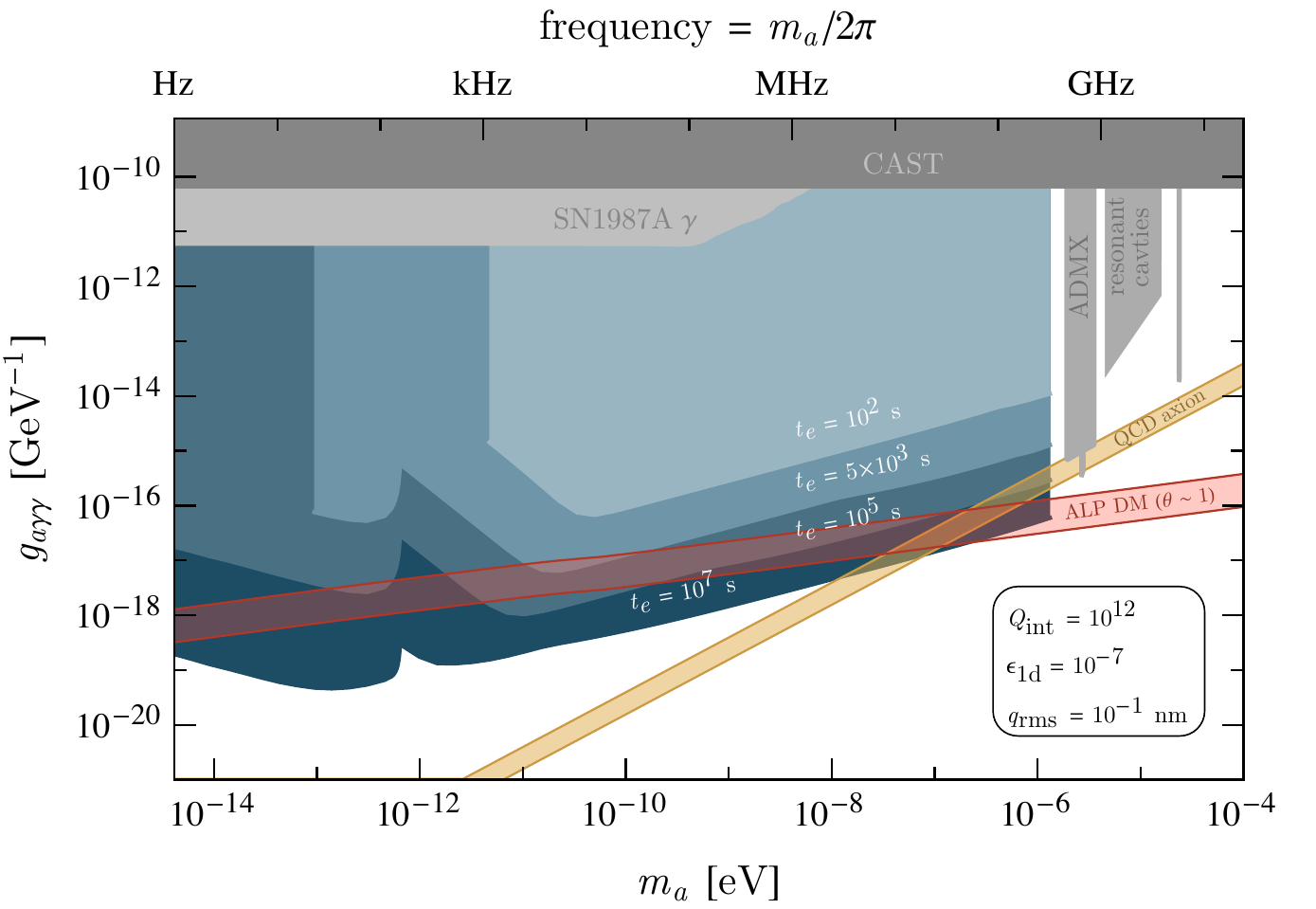}
\caption{The anticipated reach to axion dark matter in the $\gyy-m_a$ plane, for a wide range of experimental parameters. Our baseline parameters are those of the lower curve of Figure~\ref{fig:genreach}, including $Q_\text{int} = 10^{12}$, $\eps_{1 \text{d}} = 10^{-7}$, $q_\text{rms} = 10^{-1} \text{ nm}$, and $t_e = 10^7 \text{ s}$. In each panel, we vary one of these four parameters, while keeping the others fixed. All other features of the figures are as explained in Figure~\ref{fig:genreach}. Throughout, we only consider axion masses for which the integration time for a single scan step $t_\text{int}$ is larger than the axion coherence time and cavity ring-up time, and axion masses that are greater than the typical frequency shift due to mechanical vibrations.
}
\label{fig:four_plot}
\end{figure}

With our noise estimates in place, we now compute the conceptual reach of our setup. The signal PSD for the readout is slightly modified from that of Eq.~\eqref{eq:SignalPSD} because the readout receives a fraction $Q_1 / Q_{\text{cpl}}$ of the power delivered to the cavity, where $Q_1$ and $Q_\text{cpl}$ are related by Eq.~(\ref{eq:Qsum}). Referring to Eq.~(\ref{eq:SignalPSDnarrow}), we therefore make the replacement
\be
\label{eq:signal_readout_PSD}
S_\text{sig} (\w) \to \frac{Q_1}{Q_\text{cpl}} \, S_\text{sig} (\w)
\, .
\ee
This is to be compared to the total noise PSD for the readout, 
\be
\label{eq:total_noise_PSD}
S_\text{noise}(\w) = S_{\text{ql}}(\w) + \frac{Q_1}{Q_{\text{cpl}}} \left(S_{\text{th}}(\w) + S_{\text{phase}}(\w) + S_{\text{mech}}^{(1)}(\w)\right)+\frac{Q_0}{Q_{\text{cpl}}} \, S_{\text{mech}}^{(0)}(\w)\, ,
\ee
where we do not include field emission noise (see Eq.~(\ref{eq:fieldPSD})) because it can be kept below thermal noise for cylindrical cavities. Amplifier noise does not receive a factor of $Q_1 / Q_{\text{cpl}}$ because it is intrinsic to the amplifier itself. The last term in Eq.~(\ref{eq:total_noise_PSD}) corresponds to the pump mode contribution to mechanical noise (see Eq.~(\ref{eq:MechPSD})), which is rescaled by $Q_0 / Q_\text{cpl}$ since it arises from the pump mode readout coupling.

At this point, one can see why it can be advantageous to overcouple: referring to Eqs.~\eqref{eq:thermal} and \eqref{eq:Sphase}, the signal, thermal noise, and phase noise PSDs are all proportional to $Q_1^2 / Q_{\text{cpl}}$. Therefore, if either of these noise sources dominates, overcoupling ($Q_\text{cpl} \simeq Q_1 \ll Q_\text{int}$) preserves the ratio $S_{\text{sig}}(\w) / S_\text{noise}(\w)$ but broadens the frequency range that a scan step is sensitive to, relative to critical coupling ($Q_\text{cpl} = Q_\text{int}$).

We now describe the scan optimization. For a scan step with integration time $t_{\text{int}}$, the noise power is independent between frequency bins of width $\sim 1/t_{\text{int}}$. Each bin thus has an independent SNR, and the bins may be combined by a weighted average. The optimal weighting leads to an overall SNR that is the sum of the SNRs of the bins in quadrature~\cite{Chaudhuri:2018rqn}. As in Section~\ref{sec:signal}, we assume that $t_{\text{int}} > \max(\tau_{\text{r}}, \tau_a)$ and hence that the variation of $S_{\text{sig}}(\w)/S_\text{noise}(\w)$ is on frequency scales greater than $1/t_{\text{int}}$. As a result, the SNR can be approximated as an integral over frequency,
\be 
\label{eq:dicke_integral_reach}
(\text{SNR})^2 \simeq t_{\text{int}} \int_0^\infty d \w \, \left(\frac{S_{\text{sig}}(\w)}{S_\text{noise}(\w)}\right)^2
~,
\ee
where only positive frequencies are included, since the signal and noise PSDs are symmetric in $\w$.

We assume that a scan is performed uniformly in $\log{m_a}$, allocating a time $t_e$ for each $e$-fold in axion mass. It is optimal to scan in steps as wide as possible, as time must be spent waiting for the signal to ring up during each step. We take the width $\Delta \w_{\text{sc}}$ of a single scan step to be set by the range of axion masses near $\w_1 - \w_0$ within which the expected SNR, as given by Eq.~\eqref{eq:dicke_integral_reach}, is within an $\order{1}$ factor of the maximal value. Parametrically, this implies
\be
\Delta \w_{\text{sc}} \sim \max(m_a / Q_a \, , \, \w_1 / Q_1)
\, .
\ee
This step size in turn sets the integration time allowed for each scan step to be
\be
\label{eq:integration_time}
t_{\text{int}} \simeq t_e \, \frac{\Delta \w_{\text{sc}}}{m_a}
\, .
\ee
For each scan step, we numerically optimize the SNR as given by Eq.~\eqref{eq:dicke_integral_reach} with respect to the coupling $Q_{\text{cpl}}$, subject to the constraint $t_{\text{int}} > \max(\tau_{\text{r}}, \tau_a)$, and determine the reach by demanding $\text{SNR} \gtrsim 1$. As discussed in Section~\ref{sec:signal}, we model the axion PSD as following a virialized Maxwellian distribution. 

For concreteness, consider the case where thermal noise dominates and the next most important contribution is amplifier noise; this occurs at the largest axion masses shown in Figures~\ref{fig:genreach}, \ref{fig:NoiseSources}, and \ref{fig:four_plot}. As discussed further in Appendix~\ref{sec:overcouple}, Eq.~\eqref{eq:dicke_integral_reach} reduces to the usual Dicke radiometer equation~\cite{Dicke:1946glx}, and it is optimal to overcouple until the thermal noise is reduced to the quantum noise floor. This requires setting $Q_\text{cpl} \sim Q_\text{int} / n_{\text{occ}}$ where $n_\text{occ} \sim T / \w_1$ is the thermal occupation number, which leads to an enhancement of the SNR by a factor of $n_{\text{occ}}^{1/2}$ relative to critical coupling. In this case, the SNR is then, parametrically, 
\begin{equation}
\label{eq:simpleSNR}
\text{SNR} \sim \frac{\rhodm \, V}{m_a \, \w_1} \left( \gyy \,\eta_{10} \, B_0  \right)^2 \, \left(\frac{Q_a \, Q_\text{int} \, t_e}{T}\right)^{1/2} \, .
\end{equation}
For comparison, a similar analysis applied to a static-field LC resonator yields
\begin{equation}
\label{eq:lcSNR}
\text{SNR}^{(\text{LC})} \sim \rhodm \, V^{5/3} \left( \gyy \, B_{\text{LC}}  \right)^2 \, \left(\frac{Q_a \, Q_{\text{LC}} \, t_e}{T_{\text{LC}}}\right)^{1/2} \, .
\end{equation}
Our setup benefits from a large intrinsic quality factor $Q_{\text{int}}$ because it reduces dissipation in the cavity and hence thermal noise. As a result, for both our setup and LC resonators, the SNR continues to increase with $Q_{\text{int}}$ even after the signal power saturates, in agreement with the conclusions of Ref.~\cite{Chaudhuri:2018rqn}. However, for a fixed operational temperature and $e$-fold time, it is not useful to increase the intrinsic quality factor to arbitrarily large values, as there will be insufficient time to fully ring up the signal. For our choice of $t_e \sim 10^7 \text{ s}$ and $T \sim \text{K}$, this occurs when $Q_\text{int} \sim 10^{13}$. We also note that the SNR for our setup in Eq.~\eqref{eq:simpleSNR} is in principle valid for $m_a \gtrsim \text{GHz}$. However, in this regime we have no parametric advantage over existing cavity haloscopes, and accordingly our reach falls off rapidly due to the factor of $\omega_1 \simeq \omega_0 + m_a$ in the denominator.

The optimized reach is shown in Figure~\ref{fig:genreach} for two baselines choices of experimental parameters and in Figure~\ref{fig:four_plot} for a larger set of variations. In both figures, we also show existing exclusions from cavity haloscopes~\cite{Du:2018uak, PhysRevD.69.011101, Brubaker:2016ktl, PhysRevLett.59.839,Wuensch:1989sa,Hagmann:1990tj, McAllister:2017lkb}, helioscopes~\cite{Anastassopoulos:2017ftl, PhysRevLett.107.261302, PhysRevLett.112.091302, Arik:2015cjv}, and observations of SN1987A~\cite{Raffelt:1996wa,Payez:2014xsa}. We highlight parameter space that is motivated by the QCD axion as a solution to the strong CP problem, corresponding to a range bounded by the DFSZ~\cite{Dine:1982ah, Dine:1981rt} and KSVZ~\cite{Zhitnitsky:1980tq,Kim:1979if,Shifman:1979if} models. We also highlight regions of parameter space where an axion-like particle acquires a cosmological abundance from the misalignment mechanism that is in agreement with the observed DM energy density. This occurs if the initial field amplitude at the onset of oscillation is $a_i \sim (T_\text{eq}^2 m_\text{pl}^3 / m_a)^{1/4}$, where $T_\text{eq} \sim \text{eV}$ is the temperature at matter-radiation equality and $m_\text{pl}$ is the Planck mass. Defining the initial misalignment angle as $\theta_i \equiv a_i / f_a$ and relating the axion decay constant to its photon coupling by $\gyy \sim \alpha_\text{em} / 2 \pi f_a$, we find 
\be
\gyy \sim \frac{\alpha_\text{em}}{2 \pi} \,  \bigg(\frac{m_a}{m_\text{pl}^3 \, T_\text{eq}^2} \bigg)^{1/4} \, \theta_i \sim 10^{-16} \text{ GeV}^{-1} \left( \frac{m_a}{\mu \text{eV}} \right)^{1/4} \, \theta_i
\, .
\ee
This relation, along with the parametric expectation $\theta_i \sim 1$, provides a cosmologically motivated target for axion-like particles.

The projected reach covers unexplored parameter space relevant for the QCD axion for $\unit[10^{-8}]{eV} \lesssim m_a \lesssim \unit[10^{-6}]{eV}$ and for axion-like particles as light as $m_a \sim \unit[10^{-14}]{eV}$. The $m_a$ dependence of the projected sensitivity can be understood as follows. Consider the upper curve of Figure~\ref{fig:genreach}, which displays all of the parametric regimes. At large axion masses, thermal noise dominates and the reach in coupling grows as $1/\sqrt{m_a}$ when $m_a$ is decreased, as shown in Eq.~\eqref{eq:simpleSNR}. For $m_a\lesssim 10^{-10}$~eV, oscillator phase noise becomes the dominant background. The reach changes slope, degrading at smaller axion masses because the signal frequency is closer to the pump mode frequency, where the pump mode power is concentrated.  At even smaller axion masses, mechanical vibrations become the dominant source of noise, accounting for the change in slope of the reach curve around $m_a\sim 10^{-11} \text{ eV}$. This is due to the rapid increase of the mechanical noise power, as shown in Eq.~\eqref{eq:Pmech}. Near $m_a \simeq \text{kHz}\, \simeq 10^{-12}$~eV, mechanical noise decreases, as discussed in Section~\ref{sec:noiseVibration}, because of the absence of mechanical resonances below $\unit[1]{kHz}$. This result is in agreement with the experimental characterization of similar cavities performed in Ref.~\cite{Bernard:2001kp}. 

This general description also applies to the four panels of Figure~\ref{fig:four_plot}, which are intended to demonstrate the robustness of our approach. The mass dependence of the reach is qualitatively similar, except that not all of the noise regimes are always realized. For instance, in the lower left panel, oscillator phase noise never dominates over mechanical noise if $q_{\text{rms}}$ is large. Figure~\ref{fig:four_plot} shows that, as long as a large intrinsic quality factor is maintained, our approach is still sensitive to the QCD axion, even if one degrades the geometric rejection factor by $10^4$, shortens the $e$-fold time by $10^2$, or increases the amplitude of the wall vibrations by $10^6$. In all cases, our approach also still has the potential to cover a wide range of parameter space motivated by axion-like particle DM. 
 
In Figures~\ref{fig:genreach} and \ref{fig:four_plot}, the projected sensitivity of our setup is not shown for $m_a \lesssim \text{Hz}$; this corresponds to the level of frequency (and frequency splitting) control with current technology~\cite{fnalex}. As discussed in Sections~\ref{sec:realisticCavity} and~\ref{sec:noiseVibration}, this corresponds to controlling the displacement of cavity walls at the $\sim0.1\text{ nm}$ level, allowing for scanning steps as small as $\sim0.1 \text{ Hz}-1\text{ Hz}$.
%\comkz{Need to comment on why this is okay if we have step sizes of down to $\unit[0.1]{Hz}$.} 
Furthermore, our calculation is not valid at these small masses where the axion oscillates less than once per ring-up time. We also refrain from considering axion masses less than $\w_1 \, (q_\text{rms} / V^{1/3})$, since sizable mechanical displacements of the cavity walls may impede the ability to scan over such axion masses in a controlled manner.

Finally, we note that in deriving our result for the signal PSD in Eq.~\eqref{eq:SignalPSDnarrow}, we treated the oscillator, and hence the pump mode magnetic field, as monochromatic. Since the axion effective current in Eq.~\eqref{eq:maxwell} scales as $J_\text{eff} \propto B \, \partial_t a$, the oscillator width is negligible so long as it is smaller than the axion width $m_a / Q_a$. In practice, the oscillator width is quantified by the Allan variance~\cite{rubiola2009phase}. The phase noise discussed in Section~\ref{sec:leakage} is better suited to describing the tails of the spectrum. From the manufacturer data sheet of a commercially available oscillator in Ref.~\cite{datasheet}, we conclude that the pump mode width is negligible for $m_a \gtrsim \text{kHz}$. In a setup where the oscillator can be efficiently coupled to a precise reference clock (e.g., NIST~\cite{Heavner_2014}), the pump mode width is negligible in all of the parameter space we consider.

\section{Outlook}\label{sec:outlook}
%!TEX root = axion_arxiv.tex

In this work, we have proposed an approach to leverage the properties of SRF cavities to detect low-mass axions, with sensitivity to significant new parameter space spanning eight orders of magnitude in axion mass. As shown in Eq.~\eqref{eq:SNR_comparison} and confirmed in Eq.~\eqref{eq:simpleSNR}, our frequency conversion approach is parametrically enhanced compared to static-field LC resonators because of the larger electromotive forces attained (see Eq.~\eqref{eq:conceptualemf}). In addition, the insights of Ref.~\cite{Chaudhuri:2018rqn}, which shows that the sensitivity is optimized for a strongly overcoupled readout ($Q_1 \ll Q_{\text{int}}$), allow us to take advantage of the extremely large intrinsic quality factors of SRF cavities within a reasonable scanning time. 

Estimating the sensitivity of our approach required careful consideration of several noise sources. Aside from thermal noise, which can be treated relatively straightforwardly, we have pinned all of our noise estimates to quantities measured in real apparatuses. As such, we are indebted to the decades of work done on the development of quantum noise limited amplifiers, SRF cavity fabrication and testing, low phase noise oscillators, and previous precision experiments targeting both axions and gravitational waves. Ultimately, we find that for our reference parameters thermal noise is expected to dominate over most of the mass range, with vibrational noise and oscillator phase noise becoming more important at smaller masses.

We have left the detailed design of the experimental apparatus to future work. As mentioned in Sections~\ref{sec:realisticCavity} and~\ref{sec:noiseDarkCurrent}, there may be a tradeoff between maximizing the scanning range of a single cavity, and maintaining large quality factors and suppressing field emission. However, we note that even a simple cylindrical cavity design without tuning fins can potentially cover six orders of magnitude in axion mass.

In principle, our approach is also sensitive to sub-Hz axion masses. In fact, as discussed in Section~\ref{sec:signal}, the signal power is not parametrically suppressed even when the axion does not undergo a full oscillation within a resonator ring-up time. This leads to the intriguing possibility of probing axion-like particles with frequencies down to mHz or even lower. In this regime, finer details involving the stabilization of the cavity modes and the width of the oscillator become relevant, and we defer a detailed analysis to future work. 

Elaborations on our basic approach could be used to further enhance the sensitivity. For example, one could use correlations between two signal modes above and below that of the pump mode ($\w_\pm \simeq \w_0 \pm m_a$) to help distinguish the signal from noise, or use several signal modes simultaneously to accelerate the scanning. When mechanical noise dominates, two cavities with distinct mechanical resonant frequencies could be used to avoid gaps in the reach. Furthermore, variations on our approach could be sensitive to other models of ultralight bosonic dark matter, such as dilaton-like scalars that couple to the mechanical modes of the cavity. By leveraging technologies developed and proven over the past few decades, our proposal is potentially sensitive to symmetry breaking scales of up to $f_a \sim \unit[10^{16}]{GeV}$, and thereby some of the highest fundamental energy scales in nature.

\vspace{5mm}
\emph{Note added: While this study was ongoing, we became aware of Ref.}~\cite{Robert}\emph{, which discusses similar ideas for axion detection.}
 
\begin{acknowledgments}
We thank Masha Baryakhtar, Saptarshi Chaudhuri, Aaron Chou, Peter Graham, Anson Hook, Junwu Huang, Yonatan Kahn, Robert Lasenby, Steve Ritz, and Jesse Thaler for valuable discussions. AB is supported by the James Arthur Fellowship. SE, PS, and NT are supported by the U.S. Department of Energy under Contract No. DE-AC02-76SF00515. SE is also supported by the Swiss National Science Foundation, SNF project number P400P2$\_$186678. KZ is supported by the NSF GRFP under grant DGE-1656518. RTD thanks MIAPP and KITP for their hospitality and support. SE thanks the Galileo Galilei Institute (GGI) for Theoretical Physics for its hospitality and support.
\end{acknowledgments}

\appendix

\section{Cavity Geometry and Overlap Factor}\label{sec:app}
%!TEX root = axion_arxiv.tex

\subsection*{Cylindrical Cavities}

The normal modes of a cylindrical cavity are grouped into TE and TM modes. We begin by reviewing facts about these modes, following the treatment in Ref.~\cite{jackson_classical_1999}. The TM modes are defined by the vanishing of the transverse electric field $\Evec_{T}$ at $z=0$ and $z=L$, where $L$ is the height of the cylinder. Thus the $z$ component of a TM mode is defined by
\beq
E_z = \psi(r, \p) \cos\left(\frac{p \pi z}{L}\right) \, ,
\eeq
for a nonnegative integer $p$. The function $\psi$ vanishes at the boundaries and obeys a transverse wave equation, and hence has solutions of the form
\beq
 \psi(r, \p) = E_0 J_m(\gamma_{mn} r) e^{i m \p} \, ,
\eeq
where $\gamma_{mn} = x_{mn} / R$, with $x_{mn}$ being the $n$th zero of the $m$th order Bessel function $J_m(x)$, and $R$ being the cylinder radius. The transverse electric and magnetic field components of a TM mode are then given by
\begin{align}
\Evec_T &= - \frac{p \pi}{L \gamma_{mn}^2} \sin \left(\frac{p \pi z}{L}\right) \boldsymbol{\nabla}_T \psi(r, \p) \, , \\
\Bvec_T &= \frac{i \epsilon \mu \omega_{mnp}}{\gamma_{mn}^2}\cos\left(\frac{p \pi z}{L}\right)~ \mathbf{\hat{z}} \times \boldsymbol{\nabla}_T \psi(r, \p) \, ,
\end{align}
where $\boldsymbol{\nabla}_T$ is the transverse part of the gradient, and $\mu \epsilon \omega_{mnp}^2 = \gamma^2_{mn} + (p \pi / L)^2$ defines the frequency of the TM$_{mnp}$ mode.

For TE modes, the boundary condition $B_z = 0$ at $z=0$ and $z=L$ impose
\beq
B_z = \phi(r, \p) \sin\left(\frac{p \pi z}{L}\right) \, ,
\eeq
for a positive integer $p$. The function $\phi$ now obeys the boundary condition $\partial H_z / \partial r|_{r=R} = 0$. Here, the solutions to the transverse wave equation are
\beq
 \phi(r, \p) = \mu B_0 J_m(\gamma'_{mn} r) e^{i m \p} \, ,
\eeq
where $\gamma'_{mn} = x'_{mn} / R$, with $x'_{mn}$ being the $n$th root of $J'_m(x)$. The transverse electric and magnetic field components of a TE mode are then given by
\begin{align}
\Evec_T &= -\frac{i \omega_{mnp}}{\mu{\gamma'_{mn}}^2} \sin \left(\frac{p \pi z}{L}\right)~ \mathbf{\hat{z}} \times \boldsymbol{\nabla}_T \phi(r, \p) \, , \\
\Bvec_T &= \frac{p \pi}{L {\gamma'_{mn}}^2} \cos \left(\frac{p \pi z}{L}\right) \boldsymbol{\nabla}_T \phi(r, \p) \, , 
\end{align}
and $\omega_{mnp}^2 = \gamma'^2_{mn} + (p \pi / L)^2$ defines the frequency of the TE$_{mnp}$ mode.

\subsection*{Overlap Factors for Cylindrical Cavities}\label{sec:overlapints}

In this section, we compute the normalized overlap factors defined in Eq.~\eqref{eq:overlap} for transitions between cylindrical cavity modes. From this point on, we set $\epsilon = \mu = 1$ for brevity. 

We begin by deriving the selection rules quoted in Section~\ref{sec:realisticCavity}. For a geometric overlap factor between two modes indexed by $(m_0, n_0, p_0)$ and $(m_1, n_1, p_1)$, the integral over $z$ gives a factor of
\begin{align}
&\int_0^L \cos\left(\frac{p_0 \pi z}{L}\right) \sin \left(\frac{p_1 \pi z}{L}\right) d z = \frac{L p_1}{\pi(p_1^2 - p_0^2)} (1+(-1)^{p_0+p_1+1}) = \frac{L p_1}{\pi (p_1^2 - p_0^2)} \times \begin{cases} 2 & p_1 + p_0 \text{ odd} \\ 0 & p_1 + p_0 \text{ even}\end{cases}\, , 
%\\
%&\int_0^L \cos\left(\frac{p_0 \pi z}{L}\right)\cos\left(\frac{p_1 \pi z}{L}\right) =  \frac{L}{2}\left(1 + \frac{\sin(2p_1 \pi)}{2p_1 \pi} \right)\ = \begin{cases} L, ~p_1=0 \\ \frac{L}{2}, ~p_1 >0 \end{cases}, \\
%&\int_0^L \sin\left(\frac{p_0 \pi z}{L}\right)\sin\left(\frac{p_1 \pi z}{L}\right) = \frac{L}{2}\left(1 - \frac{\sin(2p_1 \pi)}{2p_1 \pi} \right) = \frac{L}{2},~~ \forall ~ p_1 \geq 0\ .
\end{align}
while the integral over $\p$ gives a factor of
\begin{align}
\int_0^{2\pi} e^{-i m_0 \p} e^{i m_1 \p} \, d\p = \begin{cases} 2\pi & m_0 = m_1 \\0 & m_0 \neq m_1 \end{cases} \, .
\end{align}
Therefore, a nonzero geometric overlap factor is only possible if the selection rules $m_0 = m_1 = m$ and $p_0 + p_1$ odd are obeyed. Furthermore, TM $\leftrightarrow$ TM mode transitions always have a zero overlap integral, because for two TM modes,
\begin{align}
\nonumber \int_V \Evec_{m n_1 p_1}^* \cdot \Bvec_{m n_0 p_0} &\propto \int_0^R d r \, r \left(\boldsymbol{\nabla}_T \psi_{m n_1 p_1}^*(r, \p) \right) \cdot \left(\mathbf{\hat{z}} \times \boldsymbol{\nabla}_T \psi_{m n_0 p_0}(r, \p) \right) \\
\nonumber &\propto \int_0^R d r \, \big( \partial_r J_{m}(\gamma_{m n_1} r ) \big) J_{m}(\gamma_{m n_0} r ) +J_{m}(\gamma_{m n_1} r ) \partial_r \big(J_{m}(\gamma_{m n_0} r )\big) \\
\nonumber &\propto \int_0^R d r \, \partial_r\big(  J_{m}(\gamma_{m n_1} r )  J_{m}(\gamma_{m n_0} r )\big) \\
\nonumber &= J_{m}(\gamma_{m n_1} R) J_{m}(\gamma_{m n_0} R) \, \\
\nonumber &= 0
\end{align}
where the last line follows from the definition of $\gamma_{mn}$. 

For TE$_{m_1 n_1 p_1}$ $\leftrightarrow$ TM$_{m_0 n_0 p_0}$ transitions, the overlap integral can be nonzero. Assuming the selection rules are obeyed, the overlap integral is
\begin{multline}
\int_V \left( \Evec^*_1\right)_{\rm TE} \cdot \left( \Bvec_0\right)_{\rm TM} = B_1 E_0 \left(\frac{\w_{m n_1 p_1}\w_{m n_0 p_0}}{(\gamma'_{m n_1})^2(\gamma_{m n_0})^2} \right) \left( \frac{4 L p_1}{p_1^2 - p_0^2}\right) \\ \times \int_0^R r \, dr \left[ \partial_r J_m (\gamma'_{m n_1} r)\partial_r J_m (\gamma_{m n_0} r) + \frac{m^2}{r^2}J_m (\gamma'_{m n_1} r) J_m (\gamma_{m n_0} r) \right] \ .
\end{multline}
The volume integral $\int_V \left( \Bvec^*_1\right)_{\rm TE} \cdot \left( \Evec_0\right)_{\rm TM}$ yields the same result, as expected. For TE$_{m_1 n_1 p_1}$ $\leftrightarrow$ TE$_{m_0 n_0 p_0}$ transitions, the overlap integral can also be nonzero. The same selection rules apply, with the additional requirement $m >0$. The overlap integral can then be written compactly as
\beq
\int_V \left( \Evec^*_1\right)_{\rm TE} \cdot \left( \Bvec_0\right)_{\rm TE} = B_1 B_0 \left(\frac{\w_{m n_1 p_1}\, m\, p_0}{(\gamma'_{m n_1})^2(\gamma'_{m n_0})^2} \right) \left( \frac{8 \pi p_1}{p_1^2 - p_0^2}\right) \Big[J_m (\gamma'_{m n_1} R) J_m (\gamma'_{m n_0} R) \Big] \, .
\eeq
To obtain the normalized overlap factor $\eta_{10}$ defined in Eq.~\eqref{eq:overlap}, one must also compute the norms of the modes,
\begin{alignat}{2}
\int_V (\Evec_1^*\cdot\Evec_1)_{\text{TE}} &= \pi L B_1^2 \ \, \frac{\w^2_{m n_1 p_1}}{(\gamma'_{m n_1})^4} \, &\int_0^R r \, dr \left[ \left(\partial_r J_m (\gamma'_{m n_1} r)\right)^2 + \frac{m^2}{r^2}\left(J_m (\gamma'_{m n_1} r)\right)^2 \right] \, , \\
\int_V (\Bvec_0^*\cdot\Bvec_0)_{\text{TE}} &= \pi L B_0^2 \, \frac{(p_0 \pi/L)^2}{{(\gamma'_{mn_0}})^4} \, &\int_0^R r \, dr \left[ \left(\partial_r J_m (\gamma'_{m n_0} r)\right)^2 + \frac{m^2}{r^2}\left(J_m (\gamma'_{m n_0} r)\right)^2 \right] \, , \\
\int_V (\Bvec_1^*\cdot\Bvec_1)_{\text{TM}} &= \pi L E_0^2 \ \, \frac{\w^2_{m n_0 p_0}}{(\gamma_{m n_1})^4} \, &\int_0^R r \, dr \left[ \left(\partial_r J_m (\gamma_{m n_0} r)\right)^2 + \frac{m^2}{r^2}\left(J_m (\gamma_{m n_0} r)\right)^2 \right] \, .
\end{alignat}
We can now write relatively compact expressions for the overlap factors,
\begin{multline}
\eta_{\rm TE\leftrightarrow TM} = \frac{4\, p_1}{\pi(p_1^2 - p_0^2)} \\ \times \frac{\int_0^R r \, dr \left[ \partial_r J_m (\gamma'_{m n_1} r)\partial_r J_m (\gamma_{m n_0} r) + \frac{m^2}{r^2}J_m (\gamma'_{m n_1} r) J_m (\gamma_{m n_0} r) \right] }{ \left(\int_0^R r \, dr \left[ \left(\partial_r J_m (\gamma'_{m n_1} r)\right)^2 + \frac{m^2}{r^2}\left(J_m (\gamma'_{m n_1} r)\right)^2 \right] \right)^{1/2}  \left(\int_0^R r \, dr \left[ \left(\partial_r J_m (\gamma_{m n_0} r)\right)^2 + \frac{m^2}{r^2}\left(J_m (\gamma_{m n_0} r)\right)^2 \right] \right)^{1/2}}
\end{multline}
and
\begin{multline}
\eta_{\rm TE\leftrightarrow TE} = \frac{8\, m\, p_1}{\pi(p_1^2 - p_0^2)} \\ \times \frac{J_m (\gamma'_{m n_1} R) J_m (\gamma'_{m n_0} R)}{ \left(\int_0^R r \, dr \left[ \left(\partial_r J_m (\gamma'_{m n_1} r)\right)^2 + \frac{m^2}{r^2}\left(J_m (\gamma'_{m n_1} r)\right)^2 \right] \right)^{1/2}  \left(\int_0^R r \, dr \left[ \left(\partial_r J_m (\gamma'_{m n_0} r)\right)^2 + \frac{m^2}{r^2}\left(J_m (\gamma'_{m n_0} r)\right)^2 \right] \right)^{1/2}} \, .
\end{multline}
Clearly, we wish to maximize $p_1$ while keeping $p_1^2 - p_0^2$ as small as possible. Therefore, good choices might include $(p_0, p_1) = (0, 1)$ or $(1, 2)$, depending on whether the relevant frequencies have a degenerate solution to perturb around.

\subsection*{Pairs of Degenerate Modes}

The axion-induced transitions we are interested in observing would be between nearly degenerate modes. Therefore, it is useful to have an analytic result for the cavity length to radius ratio that would be required to achieve degeneracy. For TM$_{m n_0 p_0}$$\leftrightarrow$TE$_{m n_1 p_1}$ transitions, we find that 
\beq
\left(\frac{L}{R} \right)^2 = \frac{\pi \, (p_1^2 - p_0^2)}{x_{m n_0}^2 - {x'}_{m n_1}^2} \, ,
\eeq
indicating that for there to be a real solution for $L/R$, then for $p_1> p_0$ we require $x'_{m n_1}<x_{m n_0}$, while for $p_1< p_0$ we require $x'_{m n_1}>x_{m n_0}$. A similar analysis can be performed for TE$_{m n_0 p_0}$$\leftrightarrow$TE$_{m n_1 p_1}$ transitions, with the same result up to a replacement of $x_{m n_0} \to x'_{m n_0}$. Tuning the length to radius ratio will then allow for axion mass to be scanned. A discussion of how tuning could be performed in practice can be found in Section~\ref{sec:realisticCavity}.

\subsection*{Corrugated Cylinders}

\begin{figure}[t]
\centering
\includegraphics[width=0.5\textwidth]{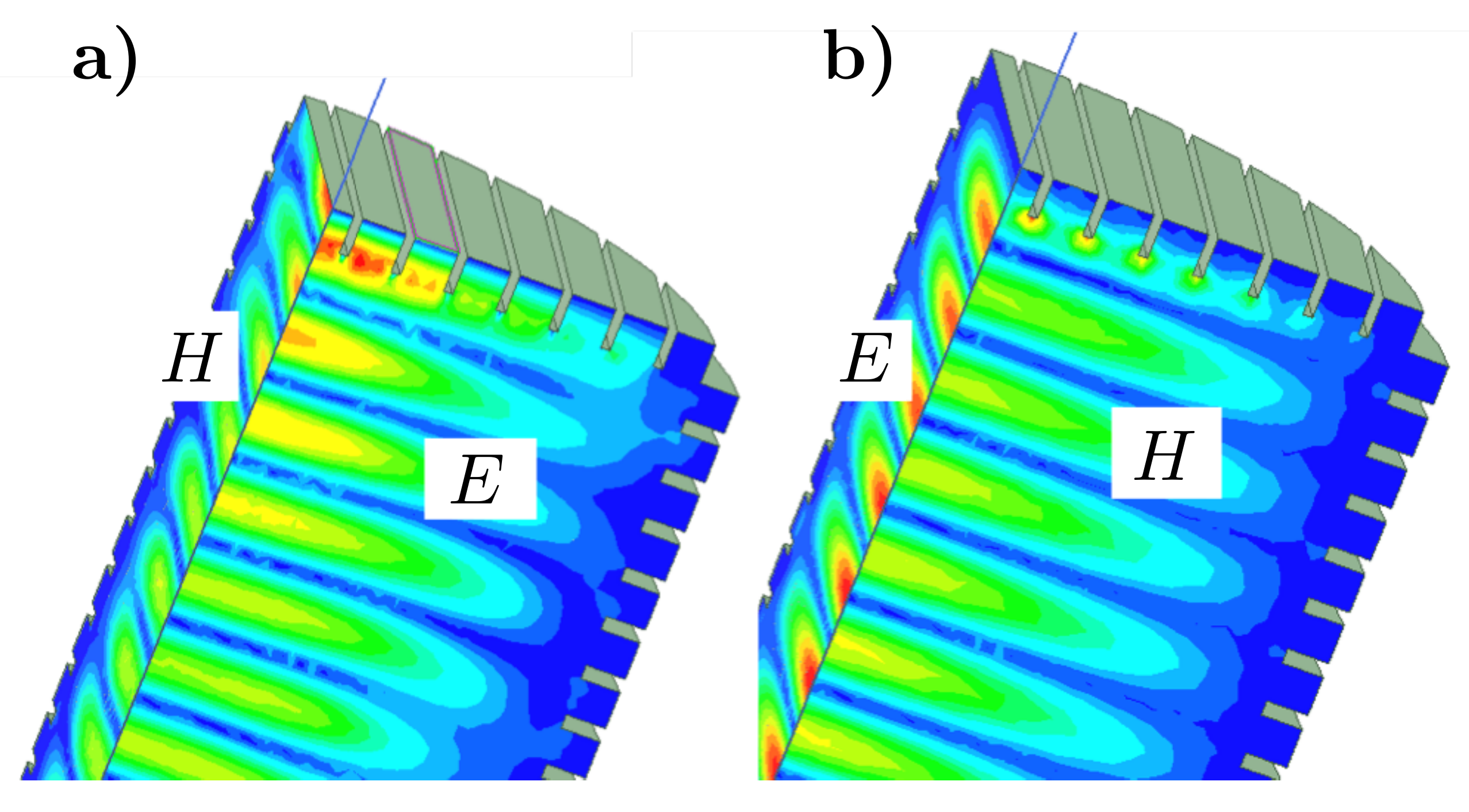}
\caption{Cavity with corrugated end- and side-walls, $R = 3 \lambda$. Shown are density plots of the $E$ and $H$ fields as labeled. \textbf{a)} Fields for mode with electric field polarized perpendicular to end-wall vanes. \textbf{b)} Fields for mode with electric field polarized parallel to end-wall vanes.}
\label{fig:corrugated}
\end{figure}

When the outer wall of a cylindrical guide can be characterized by a constant impedance $Z = E_\p / H_z$ and $Y = H_\p / E_z$, then the waveguide modes are typically hybrid, having both electric and magnetic fields transverse to the longitudinal axis $z$~\cite{5251443},
\begin{alignat}{3}
E_\p &= - \frac{k_0}{k_\perp} &\left( m \frac{k_z}{k_0} \frac{J_m (k_\perp r)}{k_\perp r} + \Lambda J_m'(k_\perp r)\right)& \sin\, m \p \, , \\
E_r &= \ \ \frac{k_0}{k_\perp} &\left( \frac{k_z}{k_0} J_m'(k_\perp r) + m \Lambda \frac{J_m (k_\perp r)}{k_\perp r}\right)& \cos\, m \p \, ,\\ 
H_\p &= \frac{k_0}{\eta_0 k_\perp} &\left( J_m'(k_\perp r) + m \frac{k_z}{k_0} \Lambda \frac{J_m (k_\perp r)}{k_\perp r}\right)& \cos\, m \p \, ,\\ 
H_r &= \frac{k_0}{\eta_0 k_\perp} &\left( m \frac{J_m (k_\perp r)}{k_\perp r} + \frac{k_z}{k_0} \Lambda J_m'(k_\perp r) \right)& \sin\, m \p \, ,
\end{alignat}
where $k_0$ and $k_z$ are the free space and longitudinal propagation constants respectively, $k_\perp = \gamma /R$ is the transverse propagation constant where $\gamma$ is a Bessel root, $\eta_0$ is the wave impedance of the medium filling the guide and $\Lambda$ is a hybrid factor relating the TE to TM fields. Imposing the boundary condition $Z = E_\p / H_z$ and $Y = H_\p / E_z$ at $r = R$ yields an equation for the hybrid factor
\beq
\Lambda = -i\left(\eta_0 Y - \frac{Z}{\eta_0} \right)\frac{k_\perp^2 R}{2 m k_z} \pm \left[1-\left(\left(\eta_0 Y - \frac{Z}{\eta_0} \right)\frac{k_\perp^2 R}{2 m k_z} \right)^2 \right]^{1/2} \, .
\eeq
The two solutions correspond to the two types of hybrid modes, HE$_{mnp}$ and EH$_{mnp}$. The most interesting case for our approach occurs when $\eta_0 Y = Z/\eta_0 \ll k_0 R$, yielding $\Lambda = \pm 1$. This limit can be obtained by using a corrugated waveguide surface and is referred to as the balanced hybrid modes. The lower order modes for this case are characterized by significantly reduced wall losses compared to the smooth wall modes. In addition, some of these modes have high degree of field polarization. The dominant balanced hybrid mode is the HE$_{11p}$, and has losses approximately 2.5 times lower than the lowest loss cylindrical mode, TE$_{01p}$, and very low cross polar fields. For a guide radius large compared to wavelength $k_z / k_0 \simeq 1$, the radial dependence of the electric and magnetic fields simplifies to $J_0(k_\perp r)$ with $\gamma = x_{10}$. For this dependence the transverse field components go to zero at the wall which explains the low attenuation. The attenuation factor $\alpha$ for the HE$_{11p}$ mode is given by 
\beq
\alpha \simeq \frac{\gamma^2}{R^3 k_0^2} \left(\frac{\omega \epsilon}{2 \sigma}\right)^{1/2} \, ,
\eeq
where it can be seen that attenuation decreases as $1/R^3$. 

The design of the detection cavity can take advantage of both of the special properties (low loss and high polarization) of the HE$_{11p}$ mode. The low wall losses allow generation of a very high $Q_{\rm int}$ cavity and the low cross-polarization coupling allows the pump and signal mode to be identical but of opposite polarization so the coupling between them is minimized. 

To achieve a high geometric overlap factor between the pump and signal modes, we again take advantage of the high polarization of the mode by introducing a polarization-dependent reflection at the cavity end walls. This can be achieved by using corrugations on the end walls as shown in Figure~\ref{fig:corrugated}. A mode with electric field polarized parallel to the corrugation vanes will be reflected at the vane edge while the mode with electric field polarized perpendicular to the vane edge will propagate into the vane section which allows for spatial alignment of the electric and magnetic fields of the two modes.

\section{Geometric Rejection}
\label{app:rejection}

\begin{figure}[t]
\centering
\includegraphics[width=0.28\textwidth]{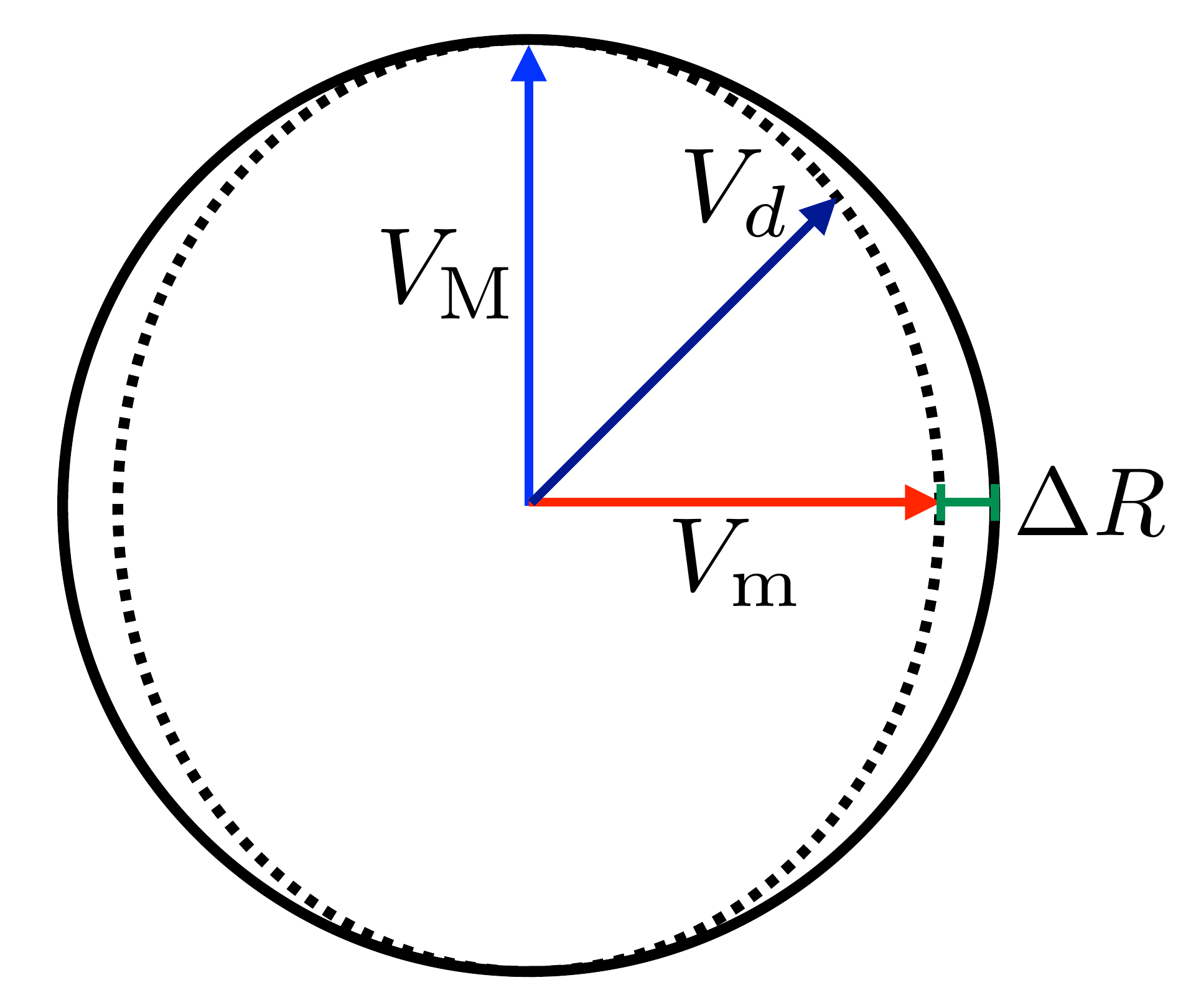}
\caption{Ellipticity of the cavity can lead to signal and pump mode contamination. Shown here are the major axis voltage $V_{\rm M}$ and minor axis voltage $V_{\rm m}$, and a drive voltage $V_{\rm d}$, subject to a wall deformation $\Delta R$.}
\label{fig:deformwall}
\end{figure}

The ability to discriminate between the pump and signal modes can be achieved by ensuring that the input (output) waveguide couples only to the pump (signal) mode to a high degree of precision. These couplings can be parametrized by geometric overlap factors, defined as
\be
\epsilon_{i j} = \frac{\int_V \tilde{\Evec}_i^* \cdot \tilde{\Evec}_j}{\sqrt{\int_V |\tilde{\Evec}_i|^2} \sqrt{\int_V |\tilde{\Evec}_j|^2}} \, ,
\ee
where the $\tilde{\Evec}_i$ denote the spatial part of the cavity normal modes as defined in Section~\ref{sec:signal}. In practice, the undesired couplings between the pump and output mode, $\epsilon_{0 \text{r}}$, and the signal and input mode, $\epsilon_{1 \text{d}}$, will be nonzero, as it is not possible to perfectly control the geometry of the various components. Below we discuss how ensuring $\epsilon_{0 \text{r}} \simeq \epsilon_{1 \text{d}} \ll 1$ can be achieved, and the required precision of the geometry of the cavity setup.

Coupling to the two orthogonal pump and detection modes in the cavity can be done through rectangular waveguides placed in the center of the two end walls rotated $90^\circ$ relative to each other. However, if there is an angular misalignment $\Delta \theta$ there will be coupling between the modes at a level proportional to the angular misalignment. For an amplitude discrimination between modes $i = 0, 1$ and drive/readout $j=$ d, r of $\epsilon_{ij} \lesssim 10^{-n}$, then $\Delta \theta \lesssim 10^{-n}$. For a guide height $h$, this would require the rotational displacement to be $\delta < h \, \Delta \theta / 2$. Assuming a frequency $f_0 = 1$ GHz, $h = \lambda /3$ and a desired power discrimination of 140 dB ($\epsilon_{ij} = 10^{-7}$), the rotational displacement error must be $\delta < 5$ nm. 

Another source of coupling of the signal and pump modes can come from small deformations of the guide resulting in ellipticity of the guide cross section, as shown in Figure~\ref{fig:deformwall}. Consider the idealized case in which the applied drive signal has a polarization midway between the major and minor axis of the elliptical guide, labeled M and m, such that it may be decomposed into the two orthogonal modes of equal amplitude. Since the propagation constant along the two elliptical axes have a slightly different value $\Delta \beta$ as a result of the radius deformation, the drive signal amplitude $V_{\rm d}(z)$ in terms of the two orthogonal modes of the elliptical guide will vary as
\beq
V_{\rm d}(z) = \frac{1}{\sqrt{2}}\left(V_{\rm M} + V_{\rm m} e^{-i \Delta \beta z} \right),~~V_{\rm M} = V_{\rm m} = V_{\rm d}(0) \, .
\eeq
This shift in relative phase between the two orthogonal modes will couple the drive mode to the readout mode as 
\beq
V_{\rm r}(z) = \frac{1}{\sqrt{2}}\left(V_{\rm M} - V_{\rm m} e^{-i \Delta \beta z} \right) = \frac{V_{\rm d}(0)}{2}\left(1 - e^{-i \Delta \beta z} \right) \, .
\eeq
The axial propagation constant of a guided mode is given by
\beq
\beta = \frac{\omega}{\left(1 - \left(k_\perp/\omega\right)^2 \right)^{1/2}} \, .
\eeq
The amplitude of the readout voltage at $z = L$ relative to the input drive voltage at $z=0$ is
\beq
\frac{V_{\rm r}(L)}{V_{\rm d} (0)} \simeq \frac{L}{2} \frac{d\beta}{dR} \Delta R = \frac{-\gamma^2 L \, \Delta R}{2 \omega R^3\left(1 - \left(k_\perp/\omega\right)^2 \right)^{3/2}} \, .
\eeq
Evaluating this expression for $f_0 = 1$ GHz, $2R = L = 5 \lambda$ and a desired power discrimination of 140 dB ($\epsilon_{1\rm d} = \epsilon_{\rm 0 r} = 10^{-7}$) we find $\Delta R \lesssim 0.2$ $\mu$m.

\section{Parametric Optimization of Coupling} \label{sec:overcouple}
%!TEX root = axion_arxiv.tex

In this section, we show analytically that overcoupling the readout can parametrically enhance the reach of our search. Our conclusions match those of Refs.~\cite{Chaudhuri:2018rqn,Chaudhuri:2019ntz}, which provide a detailed and enlightening explanation of axion search optimization in general. We will only aim for parametric estimates, as our reach is found by numerically optimizing~\eqref{eq:dicke_integral_reach}. For simplicity, we begin by considering only thermal and amplifier noise, and take the loading to be monochromatic. 

We define the dimensionless coupling strength $\xi = Q_{\text{int}}/Q_{\text{cpl}}$, giving signal and noise PSDs of
\begin{align}
S_\text{sig}(\omega_1 + \Delta \omega) &\propto \frac{\xi}{(\xi + 1)^2} \frac{S_a(\omega_1 - \omega_0 + \Delta \omega)}{1 + (\Delta \omega / \Delta \omega_r)^2} \, , \nn \\
S_\text{noise}(\omega_1 + \Delta \omega) &\propto \frac{\xi}{(\xi + 1)^2} \frac{1}{1 + (\Delta \omega / \Delta \omega_r)^2} + \frac{1}{n_{\text{occ}}} \, , \label{eq:noise_PSD_opt}
\end{align}
where we have absorbed constants to display only the dependence on $\xi$ and $\omega$, and expanded the PSDs near the positive frequency resonance $\omega \simeq \omega_1$. 

In all cases we will consider, the integrand of Eq.~\eqref{eq:dicke_integral_reach} will be roughly constant within an interval $\omega_{\text{max}} \pm \Delta \omega_s$, where we call $\Delta \omega_s$ the sensitivity width, and quickly falls off outside it. In this case, evaluating the integral roughly gives
\begin{equation} \label{eq:dicke_refined}
\text{SNR}(\xi) \approx \frac{S_\text{sig}(\omega_{\text{max}})}{S_\text{noise}(\omega_{\text{max}})} \sqrt{t_{\text{int}} \Delta \omega_s}
\end{equation}
which is the Dicke radiometer equation. Directly applying Eq.~\eqref{eq:noise_PSD_opt}, we have
\begin{equation} \label{eq:snr_frequency}
\frac{S_\text{sig}(\omega_1 + \Delta \omega)}{S_\text{noise}(\omega_1 + \Delta \omega)} \propto \frac{1}{1 + 1/n_{\text{eff}}} \frac{S_a(\omega_1 - \omega_0 + \Delta \omega)}{1 + (\Delta \omega / \Delta \omega_r)^2/(1 + n_{\text{eff}})}
\end{equation}
where $n_{\text{eff}}$ describes the ratio of thermal to amplifier noise,
\begin{equation}
n_{\text{eff}} = \frac{\xi}{(\xi + 1)^2} \, n_{\text{occ}}
\end{equation}
and $n_{\text{occ}} \gg 1$. 

We now optimize the coupling $\xi$. When the axion is broad, the sensitivity width is determined by the width of the Breit--Wigner in~\eqref{eq:snr_frequency},
\begin{equation}
\Delta \omega_s = \Delta \omega_r \sqrt{1 + n_{\text{eff}}} \propto (1 + \xi) \sqrt{1 + n_{\text{eff}}}
\end{equation}
where the second step follows because $\Delta \omega_r \propto 1/Q_1$. The maximum SNR depends on $\xi$ as 
\begin{equation}
\frac{S_\text{sig}(\omega_{\text{max}})}{S_\text{noise}(\omega_{\text{max}})} \propto \frac{1}{1 + 1/n_{\text{eff}}}.
\end{equation}
Finally, the scan step affects the SNR through the integration time, $t_{\text{int}} \propto \Delta \omega_{\text{sc}}$, as in Eq.~\eqref{eq:integration_time}, but in the broad axion case, $\Delta \omega_{\text{sc}} = \Delta \omega_a$ is independent of $\xi$. Therefore, the figure of merit to be maximized is 
\begin{equation} \label{eq:SNR_final}
\text{SNR}(\xi) \propto F(\xi) = \frac{\sqrt{(1 + \xi) \sqrt{1 + n_{\text{eff}}}}}{1 + 1/n_{\text{eff}}}.
\end{equation}
We have normalized the figure of merit so that a critically coupled readout that naively detects only the total power within the resonator width (i.e.\ taking $\xi = 1$ and artificially setting $\Delta \omega_s = \Delta \omega_r$) has $F \sim 1$. 

In the case where the axion is narrow, the roles of the scan step $\Delta \omega_{\text{sc}}$ and sensitivity width $\Delta \omega_s$ are flipped: it is now the sensitivity width that is determined by the axion width, and the scan step that is determined by the width of the Breit--Wigner. As a result, the SNR has the exact same dependence on $\xi$, so we can handle both cases at once. 

For $\xi \approx 1$, we have $n_{\text{eff}} \gg 1$, and~\eqref{eq:SNR_final} reduces to 
\begin{equation}
F(\xi) \approx \sqrt{(1 + \xi) \sqrt{n_{\text{eff}}}} = (\xi n_{\text{occ}})^{1/4}
\end{equation}
which makes it clear that overcoupling is advantageous. For $\xi \gg 1$, we can expand again to find
\begin{equation}
F(\xi) \approx \frac{\xi \sqrt{1 + n_{\text{occ}}/\xi}}{1 + \xi / n_{\text{occ}}} = \frac{(\xi n_{\text{occ}})^{1/4}}{(1 + \xi/n_{\text{occ}})^{3/4}}
\end{equation}
which is maximized when $\xi \sim n_{\text{occ}}$, at which point $F(\xi) \sim \sqrt{n_{\text{occ}}}$, justifying the claims made in Section~\ref{sec:reach}. As anticipated, the optimum is achieved when the thermal noise hits the quantum noise floor, $n_{\text{eff}} \approx 1$.

We now make some remarks on this result. First, our conclusions are not specific to thermal noise. Referring to Eq.~\eqref{eq:noise_PSD_opt}, we see that they hold for any source of noise which has a Breit--Wigner shape and the same dependence on the coupling $\xi$. In particular, this is true for oscillator phase noise. As such, the SNR gain from overcoupling is $\sqrt{n_{\text{occ}}} \sim 10$ for high axion masses, where thermal noise dominates, and grows further at low axion masses, where oscillator phase noise becomes larger than thermal noise. 

Second, we have assumed throughout that $t_{\text{int}} > \max(\tau_{\text{r}}, \tau_a)$, so that steady state solutions apply. A smaller integration time can be described by multiplying all time-dependent functions by a windowing function of width $t_{\text{int}}$, smearing their Fourier transforms over the width $1/t_{\text{int}}$. For example, in the broad axion case, the total signal power is penalized as
\begin{equation}
P_s \sim \begin{cases} t^2/\tau_{\text{r}} \tau_a & t \ll \tau_a, \\ t / \tau_{\text{r}} & \tau_a \ll t \ll \tau_{\text{r}}, \\ 1 & \tau_{\text{r}} \ll t.  \end{cases}
\end{equation}
This signal power is also smeared over a larger frequency range, so detecting it requires taking in more noise, further reducing the SNR. Therefore, to avoid dealing with excessive and unproductive casework, we have simply imposed $t_{\text{int}} > \max(\tau_{\text{r}}, \tau_a)$ as a constraint. As a result, if $t_e$ is sufficiently short (as in the lower right panel of Figure~\ref{fig:four_plot}), the readout is overcoupled beyond the optimal value to allow the scan to complete in time. In the opposite limit of large integration times, $t_{\text{int}} \gg \max(\tau_{\text{r}}, \tau_a)$, the substructure of the axion signal could be resolved, as discussed in Ref.~\cite{foster2018revealing}.

Finally, we note one more feature of our optimization: the sensitivity width $\Delta \omega_s$ can be parametrically larger than the resonator width $\Delta \omega_r$. For example, for critical coupling, we have $\Delta \omega_s \sim \sqrt{n_{\text{occ}}} \, \Delta \omega_r$. A critically coupled experiment which naively looks only at the power within the resonator width thus parametrically underestimates its potential SNR by a factor of $F(\xi = 1) \sim n_{\text{occ}}^{1/4}$. The intuition here is that the signal and thermal noise PSDs fall off resonance with exactly the same Breit--Wigner tail, so bins far beyond the resonator width can still have high SNR. However, this point is not relevant to our final result, because once the coupling is optimized, we have $\Delta \omega_s \sim \Delta \omega_r$ again.

\bibliographystyle{utphys}
\bibliography{AxionSRF}

\end{document}